\documentclass[preprint,authoryear,9pt]{elsarticle}

\usepackage[english]{babel}
\usepackage{babel}
\usepackage{fontenc}
\usepackage{graphicx}
\usepackage{amsmath,amsthm,amssymb}
\usepackage{natbib}
\usepackage{color}
\usepackage{graphicx}
\usepackage{subfigure}
\usepackage{color}
\usepackage{newlfont}
\usepackage{bigints}
\usepackage{multirow}
\usepackage{longtable} 
\usepackage{ifthen}
\usepackage{alltt}
\usepackage{enumerate}
\usepackage[text={6.25in,8.5in},centering]{geometry}
\usepackage{tikz}
\usepackage{lscape}
\newcommand{\ba}{\begin{array} }
\newcommand{\ea}{\end{array} }
\newcommand{\bae}{\begin{eqnarray}}
\newcommand{\eae}{\end{eqnarray}}
\newcommand{\bea}{\begin{eqnarray*}}
\newcommand{\eea}{\end{eqnarray*}}
\newcommand{\be}{\begin{equation}}
\newcommand{\ee}{\end{equation}}

\newcommand{\modifyb}[1]{\textcolor{blue}{#1}}

\newcommand{\modifyg}[1]{\textcolor{green}{#1}}
\newcommand{\modifyy}[1]{\textcolor{yellow}{#1}}

\newcommand{\lx}{\left(}
\newcommand{\rx}{\right)}
\newcommand{\lz}{\left[ }
\newcommand{\rz}{\right] }

\newcommand{\pr}{{\bf Proof}~~}

\def\n{\noindent}

\usepackage{tikz}
\usetikzlibrary{shapes,arrows}
\numberwithin{equation}{section}
\newtheorem{thm}{Theorem}[section]
\newtheorem{lemma}{Lemma}[section]

\newtheorem{rem}{\bf Remark}[section]
\input epsf.sty

\journal{arXiv}

\begin{document}

\begin{frontmatter}
\title{Modeling the Allee effects induced by cost of predation fear and its carry-over effects}

\author[label1]{Sourav~Kumar~Sasmal\corref{cor1}}
\ead{sourav.sasmal@gmail.com, sourav.kumar@pilani.bits-pilani.ac.in}
\author[label2]{Yasuhiro~Takeuchi}
\ead{takeuchi@gem.aoyama.ac.jp}

\cortext[cor1]{Corresponding author.}
\address[label1]{Department of Mathematics, Birla Institute of Technology and Science, Pilani, Rajasthan 333-031, India}
\address[label2]{Department of Physics and Mathematics, Aoyama Gakuin University, Kanagawa 252-5258, Japan}

\begin{abstract}
Predation driven Allee effects play an important role in the dynamics of small population, however, such predation-driven Allee effects cannot occur for the model with type $I$ functional response. It generally occurs when a generalist predator targets some specific prey. However, apart from the lethal effects of predation, there are some non-lethal effects in the presence of predator. Due to the fear of predation, positive density dependence growth may be observed at low population density, because of reduced foraging activities. Moreover, this non-lethal effect can be carried over generations. In the present manuscript, we investigate the role of predation fear and its carry-over effects in prey-predator model. First, we study the single species model in global perspective. We have shown that depending on the birth rate, our single species model describes three types of growth dynamics, namely, strong Allee dynamics, weak Allee dynamics and logistic dynamics. Then we consider the explicit dynamics of predator, with type $I$ functional response. Basic dynamical properties, as well as global stability of each equilibria have been discussed. From our analysis, we can observe that both the fear and its carry-over effects have significant role in the stability of the coexistence equilibrium, even if for the model with type $I$ functional response. The phenomenon \textit{'paradox of enrichment'} can be observed in our model, which cannot be observed in the classical prey-predator model with type $I$ functional response. However, we can see that such phenomenon can be ruled out by choosing suitable non-lethal effect parameters. Therefore, our study shows how non-lethal effects change the dynamics of a competition model, and has important biological insights, specially for the understanding of the dynamics of small populations.
\end{abstract}


\end{frontmatter}


\section{Introduction}\label{sec_introduction}
Allee effect, is a positive density-dependence phenomena, which is defined as the positive relationship between population density and \textit{per capita growth rate} $(pgr)$ at low population density \citep{Allee1931, Courchamp2008}. In contrast to the logistic dynamics, Allee effects play an important role to the extinction of small populations. There are a number of mechanisms for which Allee effects have been observed, such as, mate limitation, cooperative defense, cooperative feeding, environmental conditioning, inbreeding depression, demographic stochasticity, etc. \citep{Courchamp2008, Stephens1999b, kramer2009evidence, Dennis1989, Lewis1993}. Apart from the above mechanisms, predator-driven Allee effects also can be observed in nature. Though, predator-driven Allee effects are limited as predator population declines, prey population also declines - leading to the negative density dependence growth. It can be observed for some specific type of functional response, such as Holling type $II$ functional response \citep{Courchamp2008, Gascoigne2004, kramer2010experimental}. However, here we will investigate the occurrence of Allee effects for Holling type $I$ functional response, in the presence of non-lethal effects of predation. Due to considerable impact of Allee effects, it seeks significant attention among theoretical ecologists in several aspects, like population ecology \citep{Courchamp2008, Dennis1989, Dennis2002, sasmal2017effect, sasmal2017predator}, biological invasion \citep{Drake2004, Lewis1993, Taylor2005}, eco-epidemiology \citep{Deredec2006, Kang_MBE2013, Sasmal_Bios2013} etc. \\

Apart from the direct killing, which is widely observed in nature, many preys modify their traits in response to the predation risk. These modified traits could be related to behavior, morphology, life history of prey. To avoid predation, prey shows various anti-predator behaviors, e.g., habitat changes, reduced foraging activities, vigilance, some physiological changes, etc. \citep{cresswell2011predation}. Such effects are known as trait-mediated indirect effect, as such effect arises from predator's influence on prey traits, rather than arises from prey densities. Such non-lethal predator effects could be immediate and can influence entire prey population over entire lifetime. It can be argued that non-lethal effects are important and are needed to consider in population ecology only when it is large compared to direct density dependent effects. However, it may not be the case even when predation rate is high. Many authors suggest that such indirect effects could be equivalent or more influential compared to density effects (direct predation) \citep{creel2008relationships, cresswell2011predation, preisser2008many}. This argument was supported by the experimental data from prey-predator interaction of larval dragonfly - \textit{Anax sp.} (predator) and bullfrog tadpoles - \textit{Rana catesbeiana} (prey) \citep{peacor2001contribution}. Some recent studies showed that among such anti-predator responses, fear of predation can play an important role as direct predation effect in prey-predator models \citep{zanette2011perceived, wang2016modelling, sasmal2018population}. Due to the predation fear, scared prey forages less, as well as it leads to some stress-related physiological changes, which impact reproduction success \citep{creel2008relationships, schmitz1997behaviorally, Elliott2016experimental}. For example, birds flee from their nests in response to their predators sound as an anti-predator response \citep{creel2008relationships, cresswell2011predation}. Though such anti-predator response may be instantly beneficial as it increases the adult survival, however, as a long-term cost it reduces the reproduction rate \citep{cresswell2011predation}. Zanettee et al. \citep{zanette2011perceived} experimentally showed the $40$\% reduction in offspring production of prey (song sparrows - \textit{Melospiza melodia}) due to predation fear by providing predatory sound only and without direct killing. They showed that this reduction is due to anti-predator behavior which affects the reproduction of song sparrows. Therefore, for free living wildlife population, incorporation of non-lethal trait-mediated indirect effect by predation fear is important. Moreover, as density declines, prey individuals are more vigilant and less foraging. Therefore, such non-lethal effect could be a cause of Allee effects and increases the extinction risk of small populations \citep{clutton1999predation, mooring2004vigilance}. \\

The 'carry-over effect' originally started from recurrent measures of clinical experiments. However, recently it has been used in ecological and evolutionary aspects and can be used for broad range of situations. O'Connor et al. \citep{o2014biological} proposed the following working definition for carry-over effects: ``In an ecological context, carry-over effects occur in any situation in which an individual's previous history and experience explains their current performance in a given situation". In view of the above definition, carry-over effects are not restricted to the seasonal requirement, discrete-time scale, migration, etc. \citep{marshall2011ecological, o2011consequences}, which was previously considered. It should be considered as a more general phenomenon, which allows us to identify in broad range of situations, like, within and across life-history stages, seasons, years etc. Under this definition, life-history trade-offs and costs of reproduction can be viewed as special types of carry-over effects. Moreover, some lab experiments showed that non-lethal carry-over effects have impact in long-term population dynamics \citep{betini2013carry, betini2013density}. Carry-over effects can occur in both across multiple seasons or within a single season (e.g., transitions between physiological states within a season). Experimental evidences of carry-over effects within a single season and over short time periods are observed in insects \citep{de2005fitness}, amphibians \citep{touchon2013effects}, marine fish \citep{johnson2008combined}, and marine invertebrates \citep{marshall2011ecological}, etc. Due to the above reasons, study on ecological carry-over effects has been increasing in mathematical modeling studies \citep{norris2005carry, runge2005modeling, norris2006predicting}, as well as in empirical research \citep{norris2004tropical, inger2010carry, legagneux2012manipulating, sedinger2011carryover}. Therefore, integrating the research on carry-over effects, with potential connection between life-history trade-offs and cost of reproduction will improve our understanding of the factors affecting population dynamics in nature. \\

Betini et al. \citep{betini2013carry} introduced an experimental model system of \textit{Drosophila} to study the sequential density dependence and carry-over effects and used a simple \textit{Ricker map} with season-specific parameters. On the other hand, Elliott et al. \citet{elliott2017fear} investigated the role of fear in relation to fitness and population density by considering a prey-predator system of \textit{Drosophila melanogaster} (prey) and mantid (predator), both in breeding and non-breeding seasons. They used the experimental results to parameterize a bi-seasonal Ricker map and provided the evidence that indirect effect of predator can be a cause of Allee effect, which is very important to understand the dynamics of small population. Motivated from the above discussion, we develop and analyze continuous-time population models to investigate the cost of predation fear and its carry-over effects in prey-predator interaction with Holling type $I$ functional response. The objectives of our study are to answer the following questions: $(i)$ How these non-lethal effects (cost of predation fear and its carry-over effects) change the growth dynamics of prey population? $(ii)$ How the phenomenon \textit{'paradox of enrichment'} can be observed for our model and under which condition it can be ruled out? $(iii)$ Role of non-lethal effects on the stability of the coexistence equilibrium. $(iv)$ What is the global dynamical behavior of our proposed model? The remainder of the paper is organized as follows: In Section \ref{sec_model_development}, we formulate and analyze a single species population model with fear and its carry-over effects. We find the global dynamical behavior of our proposed single species model, as well as we show that how growth dynamics changes depending on the parameter values. In Section \ref{sec_predator-prey_type-I}, we consider the explicit dynamics of predator population, with Holling type $I$ functional response. Basic dynamical properties, as well as global stability of equilibria are provided in this section. Moreover, existence and stability of Hopf-bifurcation is also provided in this section. Some numerical simulation results are shown in the Section \ref{sec_numerical}. In Section \ref{sec_discussion}, we discuss our results and findings and provide some potential future directions. Some detailed proofs of our analytical findings are given in Section \ref{sec_proofs}.

\section{Single species model with fear and carry-over effects}\label{sec_model_development}
First, we consider prey growth follows the logistic dynamics, which can be split into three parts, birth, natural death, and death due to intra-prey competition. Thus in the absence of predator, a single species population model is given by the following ODE:
\begin{eqnarray}\label{model_logistic}
\begin{array}{ccc}
\frac{dx}{dt} = rx - d_1x - d_2x^2,
\end{array}
\end{eqnarray}
where $x$ is the density of prey, whose maximum birth rate is $r$ in the absence of predation. $d_1$ is the natural death rate and $d_2$ is the density dependent death rate of prey. Next, we consider a single species (prey) population model with a generalist predator, which is at constant density $y$. Here, we neglect the direct predation. We develop and analyze a single species population model with fear and carry-over effects which is given by
\begin{eqnarray}\label{model_single_species}
\begin{array}{ccc}
\frac{dx}{dt} = \underbrace{rx}_{\text{birth}}\underbrace{\frac{1+cx}{1+cx+fy}}_{\text{fear and carry-over effect}} - d_1x - d_2x^2 = x\underbrace{\lz \frac{r(1+cx)}{1+cx+fy} - d_1 - d_2x\rz}_{\text{per capita growth rate}} \equiv x\Theta(x).
\end{array}
\end{eqnarray}
Here $c$ is the carry-over effect parameter due to fear, which is quantified by the parameter $f$. In the Model \eqref{model_single_species}, if $f = 0$ (i.e., if there is no growth-rate reduction due to predation fear), then the model will be simply logistic dynamics \eqref{model_logistic}. If $c = 0$, then model is reduced to the single species model with only fear effect which has been studied in \citep{Wang-Zhang2011, sasmal2020dynamics}.

\begin{lemma}
The solution of system \eqref{model_single_species} is uniformly ultimately bounded in $\mathbb{R}_+$ with $$\lim_{t\rightarrow\infty} x(t) \leq \frac{r-d_1}{d_2},$$ when $r\geq d_1$.
\end{lemma}

Here, System \eqref{model_single_species} can exhibit Allee effects depending on the parameters $r$, $c$, $f$, $d_1$, $d_2$, and when $\Theta^{'}(0)>0$. The Allee effect will be weak if population of \eqref{model_single_species} persists with $\Theta(0)>0$ and strong if there exists a threshold density (Allee threshold) below which population of \eqref{model_single_species} goes to extinction and above which population persists for $\Theta(0)<0$. Moreover, when $\Theta^{'}(0)<0$ or any one of the parameters $f$ and $c$ is zero (or both) then the system \eqref{model_single_species} shows logistic dynamics. \\

System \eqref{model_single_species} always has the extinction equilibrium $x_0 = 0$. Other equilibria of the System \eqref{model_single_species} are roots of the quadratic equation
\begin{eqnarray}\label{eqn_single_species}
\begin{array}{ccc}
\Phi(x) \equiv cd_2 x^2 + \lz c(d_1-r)+d_2(1+fy)\rz x + \lz d_1-r+d_1fy\rz = 0.
\end{array}
\end{eqnarray}
We rename the coefficients of the above equation as $$\Omega_1 = c(d_1-r)+d_2(1+fy) \mbox{ and } \Omega_2 = d_1-r+d_1fy.$$ Denote the roots of the above equation by $$x_{1,2} = \frac{-\Omega_1\pm\sqrt{\Delta}}{2cd_2},$$ where
\begin{eqnarray}\label{eqn_delta_single_species}
\begin{array}{ccc}
\Delta = \Omega_1^2-4cd_2\Omega_2 = \lz c(d_1-r)+d_2(1+fy)\rz^2 - 4cd_2\lz d_1-r+d_1fy\rz.
\end{array}
\end{eqnarray}

Here, $x_1>x_2$ (when both the roots are real, i.e., when $\Delta>0$). Further we note that $\Theta(x)<0$ for any $x>0$, when $r\leq d_1$. Hence, $x(t)\rightarrow0$ as $t\rightarrow\infty$, when $r\leq d_1$. No other non-negative equilibrium exists in this case, except $x_0 = 0$, which is globally asymptotically stable. Hereafter, we assume that $r>d_1$. \\

Depending on the number of roots of the quadratic equation \eqref{eqn_single_species}, we have the following three cases:

\begin{enumerate}
\item \eqref{eqn_single_species} has no positive solution in the following two cases:
$$\mbox{(i) } \Omega_1 \geq 0 \mbox{ and } \Omega_2\geq0, ~~~~ \mbox{ (ii) } \Omega_1 < 0 \mbox{ and } \Delta < 0$$

\item \eqref{eqn_single_species} has unique positive solution in the following three cases:
$$\mbox{(i) } \Omega_2 < 0, ~~~~ \mbox{ (ii) } \Omega_1 < 0 \mbox{ and } \Omega_2 = 0, ~~~~ \mbox{ (iii) } \Omega_1 < 0 \mbox{ and } \Delta = 0.$$

\item \eqref{eqn_single_species} has two positive solutions if
$$\Omega_1 < 0, \Omega_2 > 0 \mbox{ and } \Delta > 0.$$
\end{enumerate}

Following theorem summarizes the existence and stability of all equilibria for the Model \eqref{model_single_species}.

\begin{figure}
\begin{center}
\subfigure[No positive equilibrium, $x_0$ is globally stable for $r=1.25$. Here, $d_1(1+fy) = 2$ and $d_1+\frac{d_2(1+fy)}{c} = 1.5$.]{\includegraphics[height = 50mm, width = 70mm]{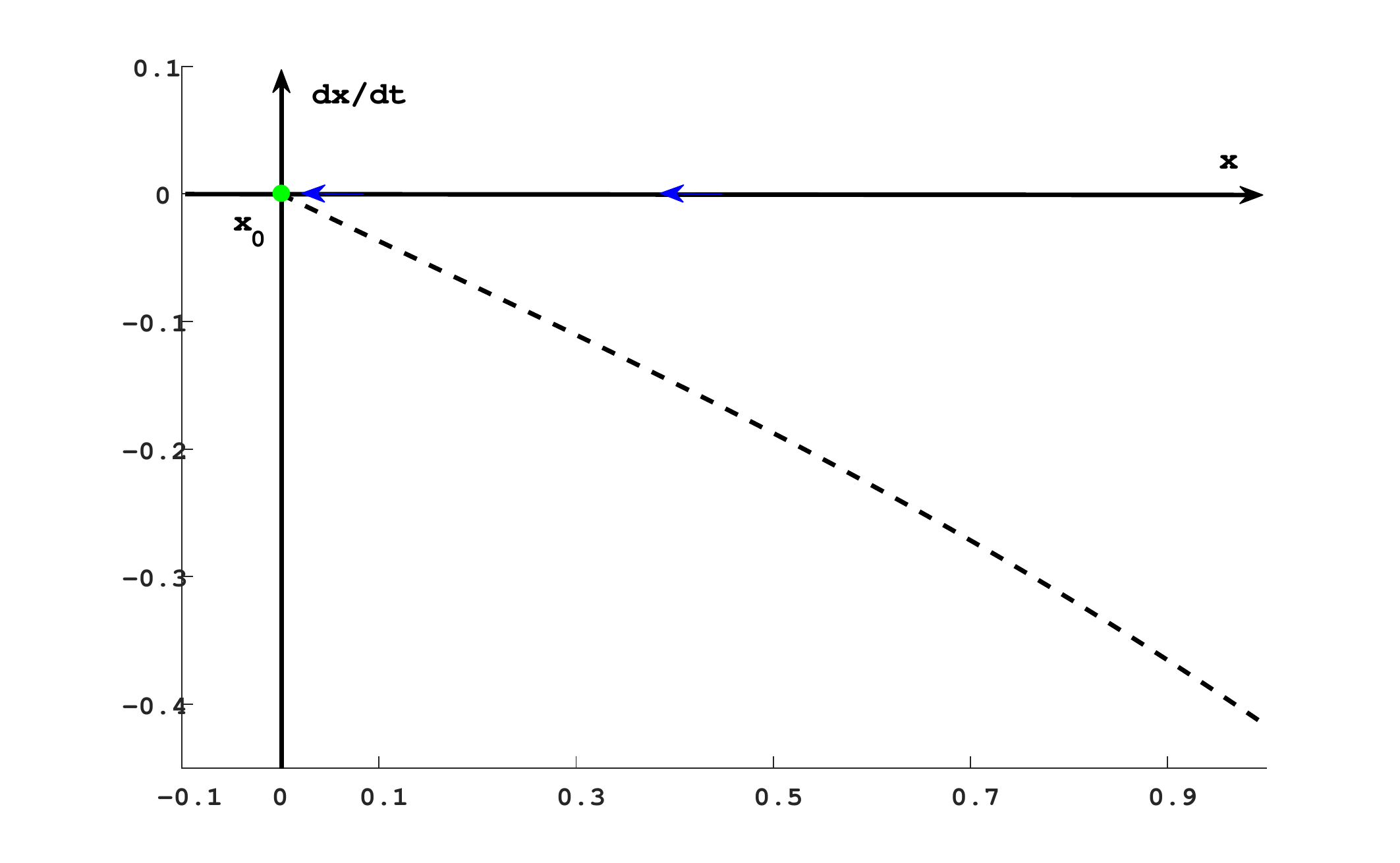}\label{fig_only_x0_1st}}\hspace{10mm}
\subfigure[No positive equilibrium, $x_0$ is globally stable for $r=1.85$. Here, $d_1+\frac{d_2(1+fy)}{c} = 1.5$ and $\Delta=-0.0275~(<0)$.]{\includegraphics[height = 50mm, width = 70mm]{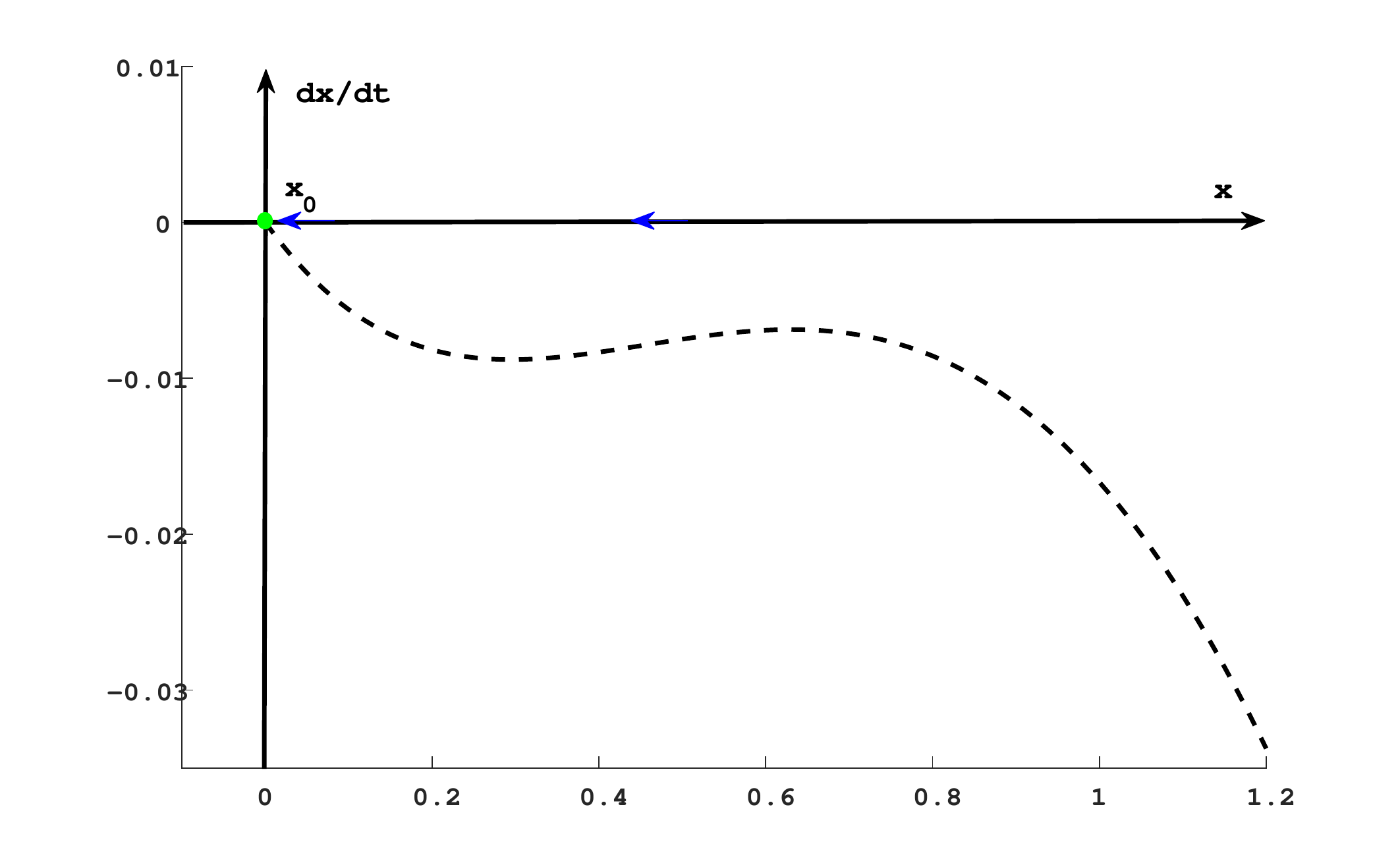}\label{fig_only_x0_2nd}}
\subfigure[Unique positive equilibrium $x_1$ is globally stable for $r=2.25$. Here, $d_1(1+fy) = 2$.]{\includegraphics[height = 50mm, width = 70mm]{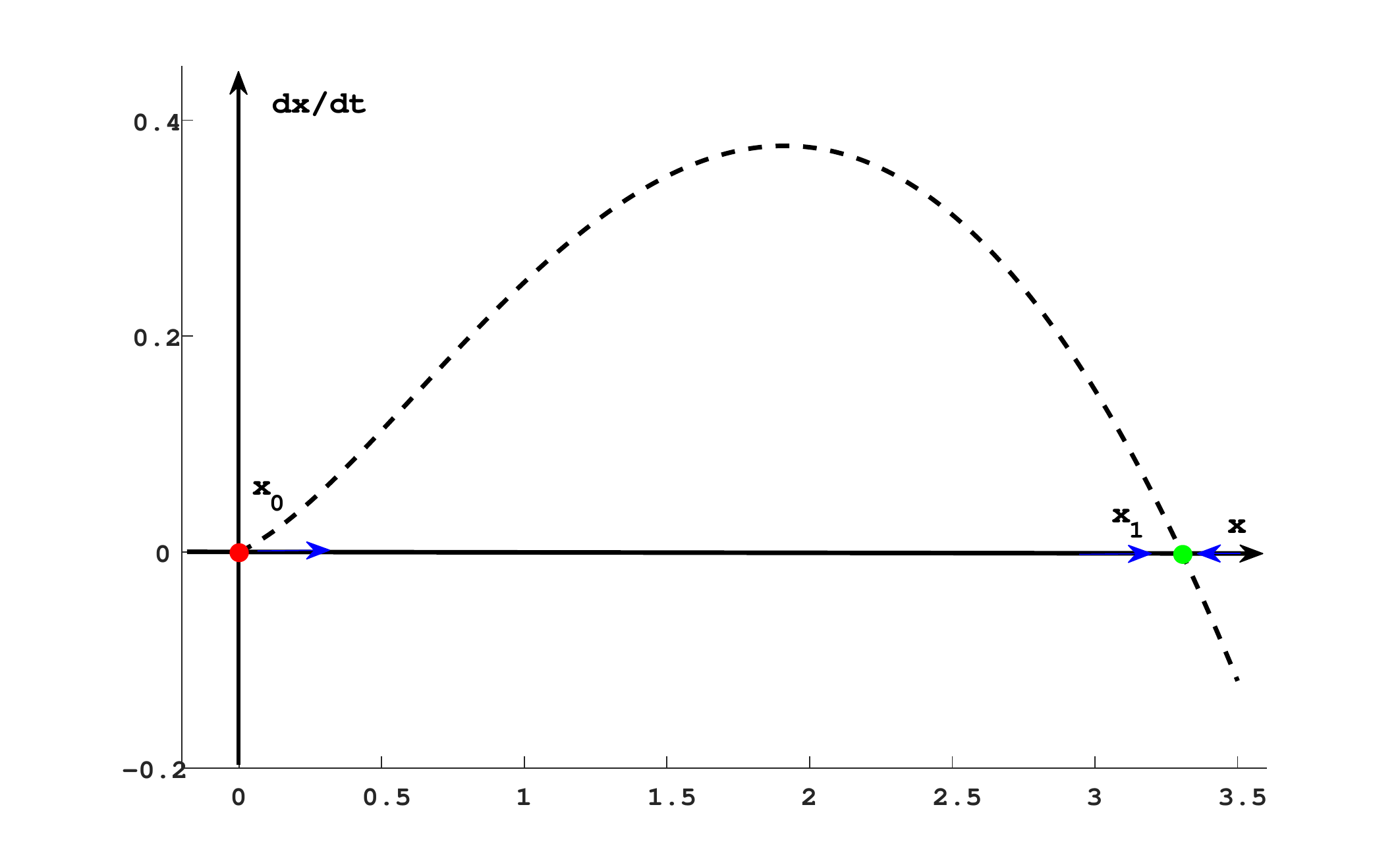}\label{fig_single_root_x1}}\hspace{10mm}
\subfigure[Unique positive equilibrium $x_1$ is globally stable for $r=d_1(1+fy)=2$. Here, $d_1+\frac{d_2(1+fy)}{c} = 1.5$.]{\includegraphics[height = 50mm, width = 70mm]{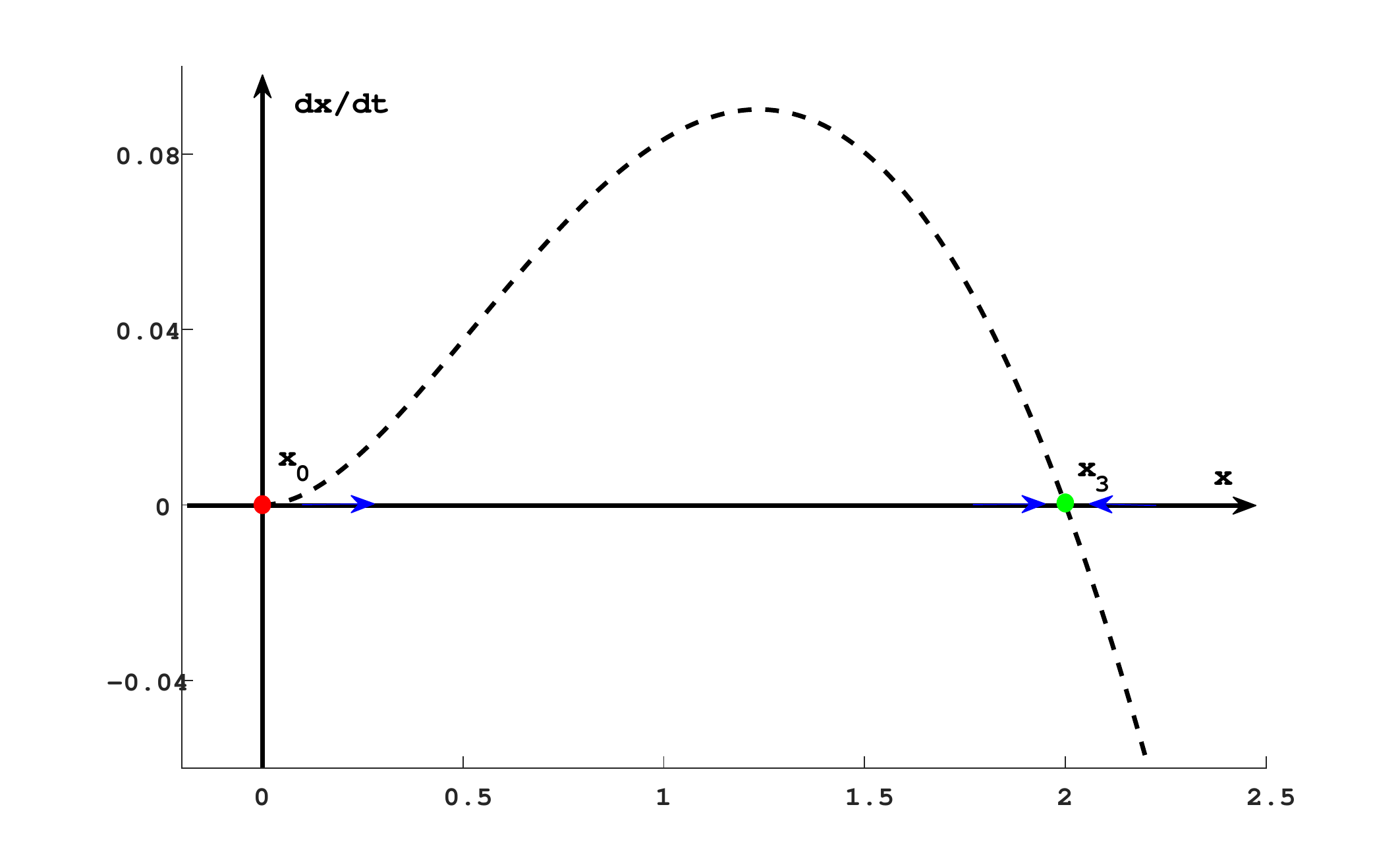}\label{fig_single_root_x3}}
\subfigure[Unique positive equilibrium $x_1$ is unstable and $x_0$ is locally stable for $r=1.866$. Here, $d_1+\frac{d_2(1+fy)}{c} = 1.5$, and $\Delta\approx0$.]{\includegraphics[height = 50mm, width = 70mm]{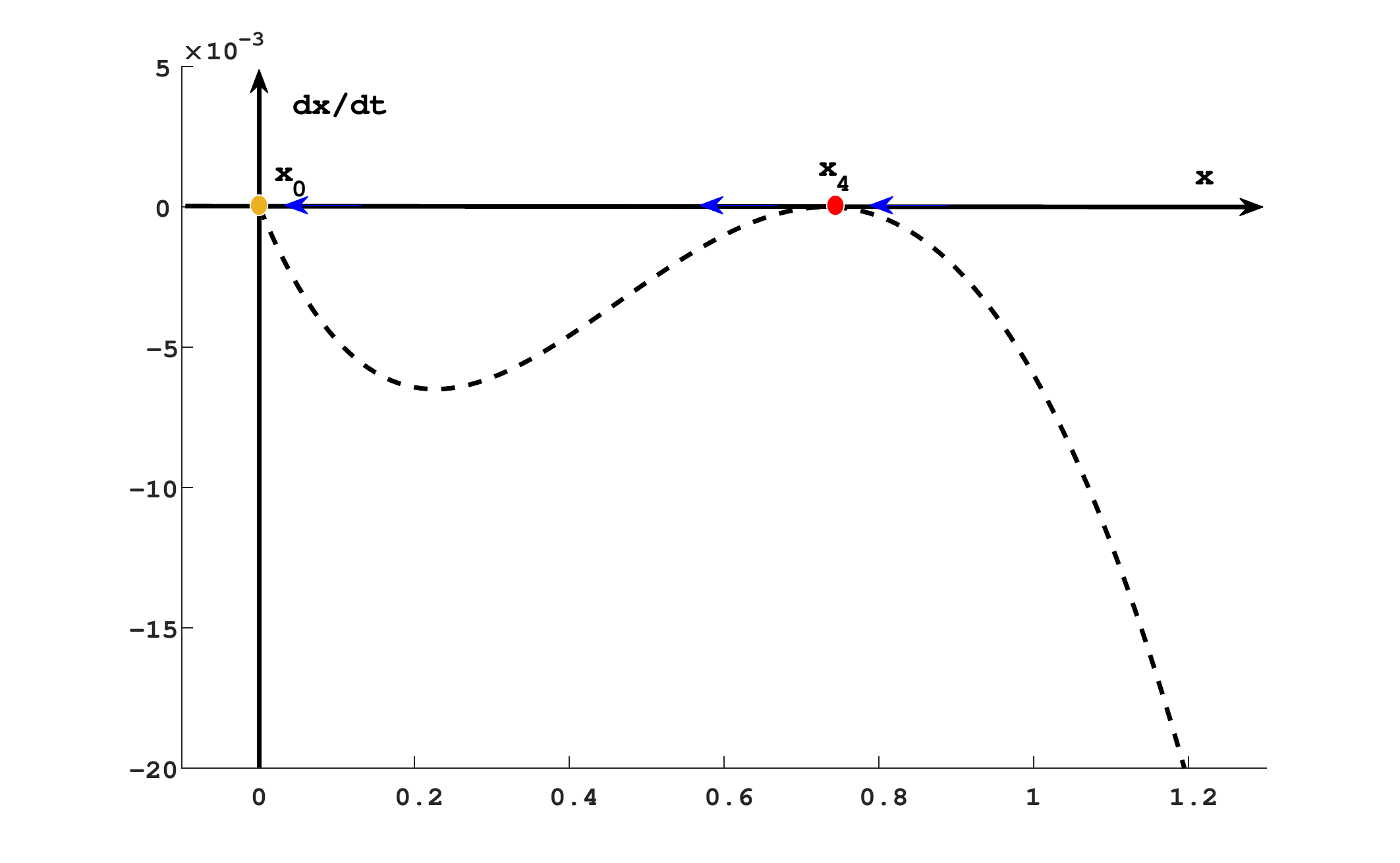}\label{fig_single_root_multiplicity_2}}\hspace{10mm}
\subfigure[Two positive equilibria $x_1$ and $x_2$ exist for $r=1.9$. $x_0$ and $x_1$ are locally stable. Here, $d_1(1+fy) = 2$, $d_1+\frac{d_2(1+fy)}{c} = 1.5$, and $\Delta=0.06~(>0)$.]{\includegraphics[height = 50mm, width = 70mm]{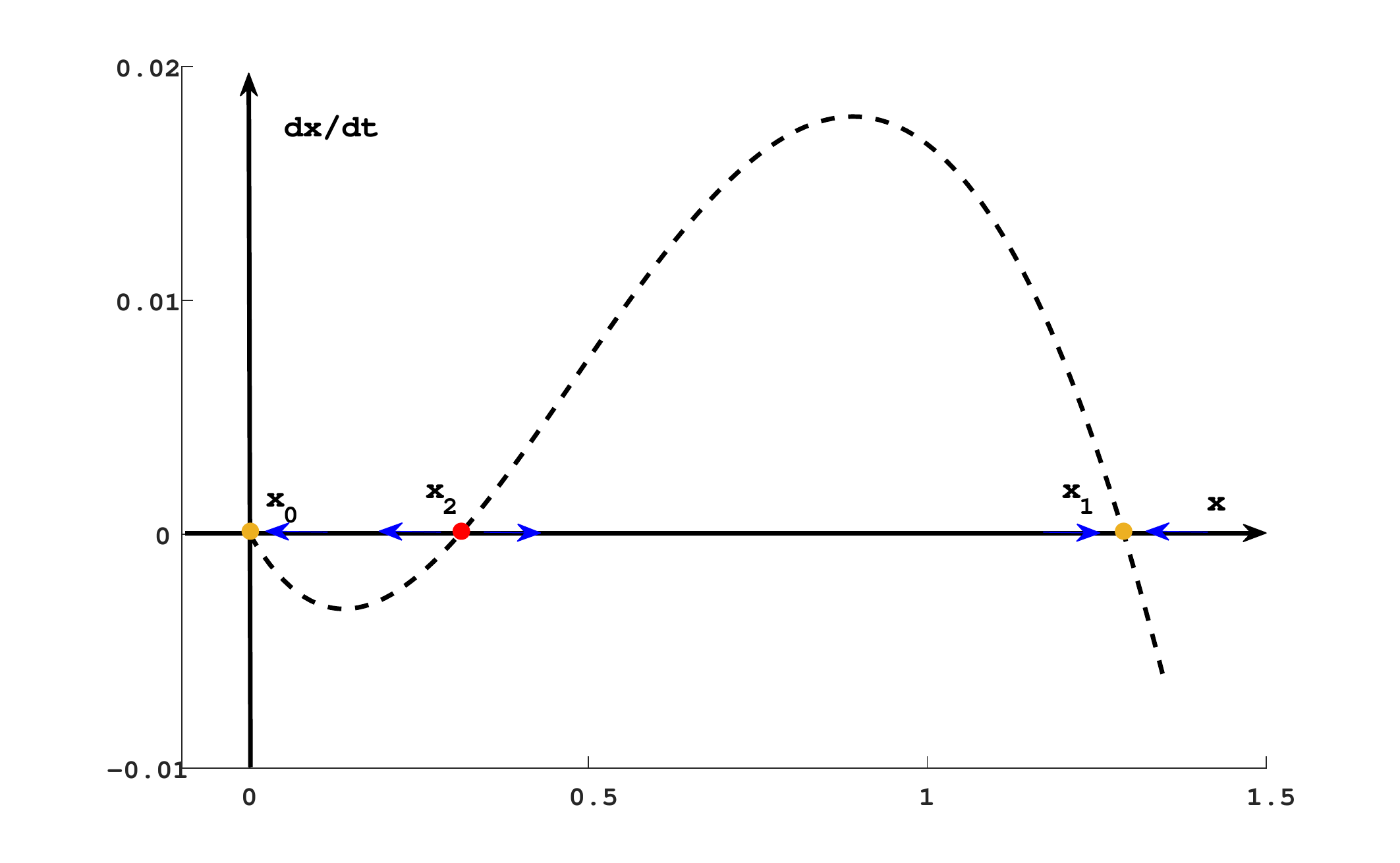}\label{fig_two_real_positive_roots}}
\end{center}
\caption{Existence of equilibria and their stability for the Model \eqref{model_single_species}. Green dots $(\modifyg{\bullet})$ are globally stable equilibria, red dots $(\bullet)$ are unstable equilibria, yellow dots $(\modifyy{\bullet})$ are locally stable equilibria, and blue arrows $(\modifyb{\bold\rightarrow})$ are the flow direction. The fixed parameters are $f=1$, $c=1$, $y=1$, $d_1=1$, $d_2=0.25$.}
\label{fig_for_thm_existence_stability_single_species}
\end{figure}

\begin{thm}\label{Theorem_existence_stability_single_species}[Existence and stability of equilibria for Model \eqref{model_single_species}.]
\begin{enumerate}
\item The System \eqref{model_single_species} has only the trivial extinction equilibrium $x_0 = 0$, if any of the following two conditions holds:
\begin{enumerate}
\item[] (i) $r\leq\min\Big\{d_1(1+fy), d_1+\frac{d_2(1+fy)}{c}\Big\}$ (see Figure \ref{fig_only_x0_1st}). \\

\item[] (ii) $r>d_1+\frac{d_2(1+fy)}{c}$ and $\Delta<0$ (see Figure \ref{fig_only_x0_2nd}).
\end{enumerate}
In this case the equilibrium $x_0=0$ is globally asymptotically stable. \\

\item The System \eqref{model_single_species} has two non-negative equilibria, namely, the trivial extinction equilibrium $x_0 = 0$, and a non-trivial equilibrium, if any of the following three conditions holds:
\begin{enumerate}
\item[] (i) If $r>d_1(1+fy)$, then \eqref{model_single_species} has unique positive equilibrium point $x_1$. Here, $x_0$ is always unstable and $x_1$ is globally asymptotically stable (see Figure \ref{fig_single_root_x1}). \\

\item[] (ii) If $r = d_1(1+fy) > d_1+\frac{d_2(1+fy)}{c}$, then \eqref{model_single_species} has unique positive equilibrium $x_3 = \frac{c(r-d_1)-d_2(1+fy)}{cd_2}$. Here also $x_0$ is unstable and $x_3$ is globally asymptotically stable (see Figure \ref{fig_single_root_x3}). \\

\item[] (iii) If $r>d_1+\frac{d_2(1+fy)}{c}$ and $\Delta=0$, then \eqref{model_single_species} has one positive equilibrium $x_4 = \frac{c(r-d_1)-d_2(1+fy)}{2cd_2}$ of order two. Here, $x_0$ is locally asymptotically stable and $x_4$ is a saddle (see Figure \ref{fig_single_root_multiplicity_2}).\\
\end{enumerate}

\item The System \eqref{model_single_species} has three non-negative equilibria, namely, the trivial extinction equilibrium $x_0$, and two non-trivial equilibria $x_1$ and $x_2$, if $d_1+\frac{d_2(1+fy)}{c} < r < d_1(1+fy)$ and $\Delta>0$. Here, both the equilibria $x_0$ and $x_1$ are locally asymptotically stable, and $x_2$ is always unstable (see Figure \ref{fig_two_real_positive_roots}).
\end{enumerate}
Where $\Delta$ is defined in \eqref{eqn_delta_single_species}.
\end{thm}

{\bf Note:} Under conditions of Theorem \eqref{Theorem_existence_stability_single_species} part $1(i)$ and $1(ii)$, pgr (per capita growth rate) of the Model \eqref{model_single_species} is always negative, ($\Theta(x) < 0$ for all $x > 0$), i.e., as a whole pgr lies below the density axis. In this case population declines and eventually goes extinct, no matter how large the population is, which is known as ``fatal Allee effects" (figures \eqref{fig_only_x0_1st} and \eqref{fig_only_x0_2nd}) \citep{Courchamp2008}. \\

In the following lemma, we discuss the possible bifurcation points for the System \eqref{model_single_species}.

\begin{lemma}\label{lemma_single_species}[Saddle-node and transcritical bifurcations.]
When the Model \eqref{model_single_species} parameters are such that $\Delta=0$, then a saddle-node bifurcation occurs at the positive equilibrium point $x_4$. Moreover, when $\Omega_2 = 0$, i.e., $r=d_1(1+fy)$, then Model \eqref{model_single_species} experiences a transcritical bifurcation at the positive equilibrium $x_3$.
\end{lemma}

\begin{figure}
\begin{center}
\includegraphics[height = 50mm, width = 120mm]{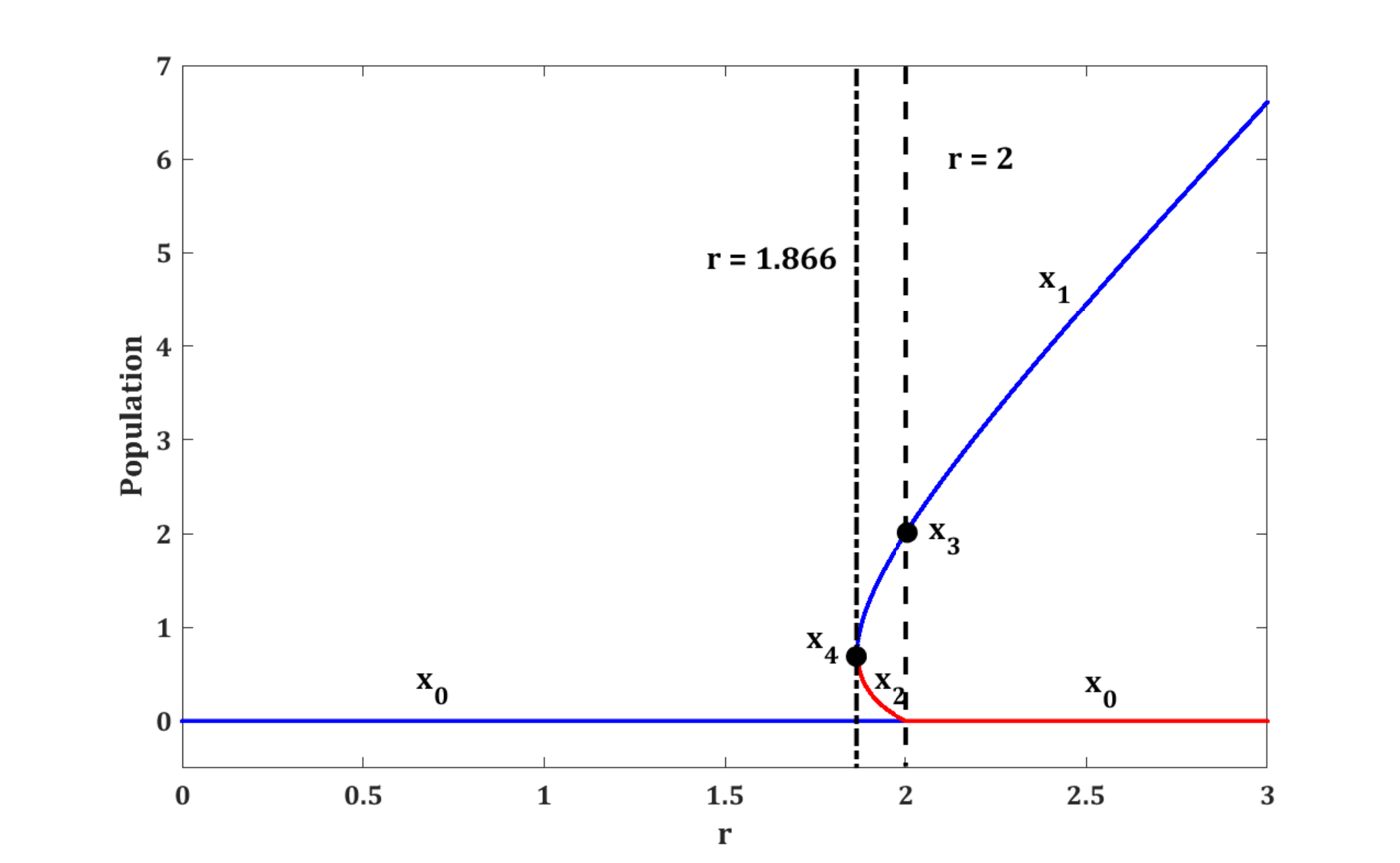}
\end{center}
\caption{Bifurcation diagram of our Model \eqref{model_single_species}, with respect to the parameter $r$. Saddle-node bifurcation occurs at $r = 1.866$ (black dot-dashed line) and transcritical bifurcation occurs at $r = 2$ (black dashed line). All the other parameters are fixed at $f=1$, $y=1$, $c=1$, $d_1=1$, and $d_2 = 0.25$.}
\label{fig_bifurcation_single_species}
\end{figure}

The saddle-node bifurcation and transcritical bifurcation have shown in the Figure \ref{fig_bifurcation_single_species}. \\

For the Model \eqref{model_single_species}, the pgr is given by $\Theta(x)$ and $\Theta^{'}(x) = \frac{rcfy}{(1+cx+fy)^2}-d_2$. Therefore, according to the previous discussion, Model \eqref{model_single_species} may show Allee dynamics if $\Theta^{'}(0)>0$, i.e., if $r>\frac{d_2(1+fy)^2}{cfy}$. Therefore, our model \eqref{model_single_species} shows three types of growth dynamics, which is summarized in the Table \eqref{table_growth_dynamics_single-species-model}.

\begin{table}
\centering
\begin{tabular}{|l|c|}
\hline
{\bf Type of growth dynamics} & {\bf Parameter constraints} \\
\hline
Weak Allee dynamics & $r>\max\Big\{d_1(1+fy), \frac{d_2(1+fy)^2}{cfy}\Big\}$ \\
&\\
Strong Allee dynamics & $\max\Big\{d_1+\frac{d_2(1+fy)}{c},\frac{d_2(1+fy)^2}{cfy}\Big\}<r<d_1(1+fy) \mbox{ and } \Delta>0$ \eqref{eqn_delta_single_species}\\
&\\
Logistic dynamics & $r<\frac{d_2(1+fy)^2}{cfy}$ or $c=0$ or $f=0$ \\
\hline
\end{tabular}
\caption{Type of growth dynamics and parameter constraints for the Model \eqref{model_single_species}.}
\label{table_growth_dynamics_single-species-model}
\end{table}

\begin{figure}
\begin{center}
\subfigure[Model \eqref{model_single_species} shows weak Allee dynamics when $r(=2.1)>\max\{d_1(1+fy),\frac{d_2(1+fy)^2}{cfy}\}(=\max\{1,2\})$ for the parameter values $r=2.1$, $f=1$ and $c=1$.]{\includegraphics[height = 50mm, width = 70mm]{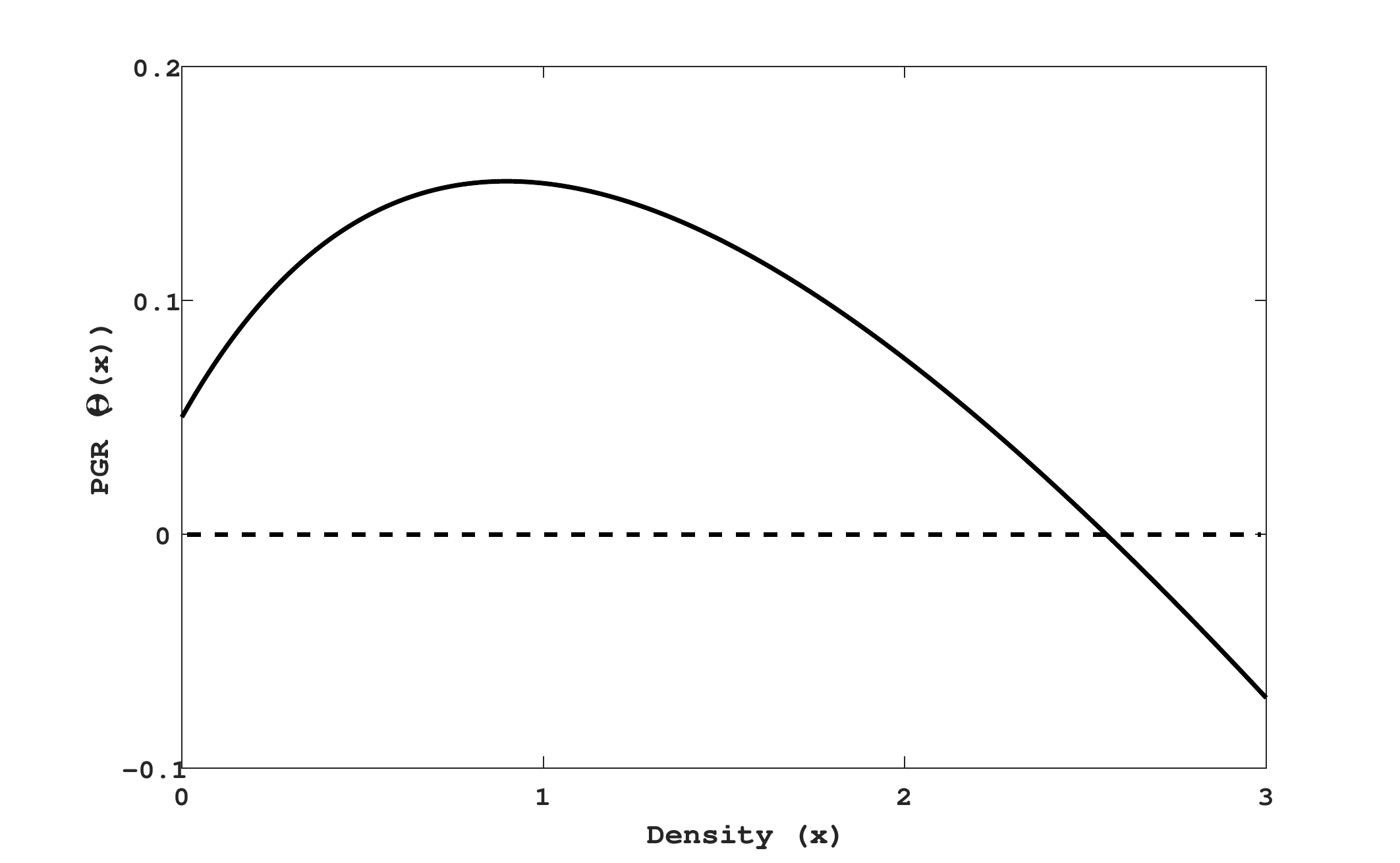}\label{fig_Weak_Allee}}\hspace{10mm}
\subfigure[Model \eqref{model_single_species} shows strong Allee dynamics when $\frac{d_2(1+fy)^2}{cfy}(=1)<r(=1.95)<d_1(1+fy)(=2)$ for the parameter values $r=1.95$, $f=1$ and $c=1$.]{\includegraphics[height = 50mm, width = 70mm]{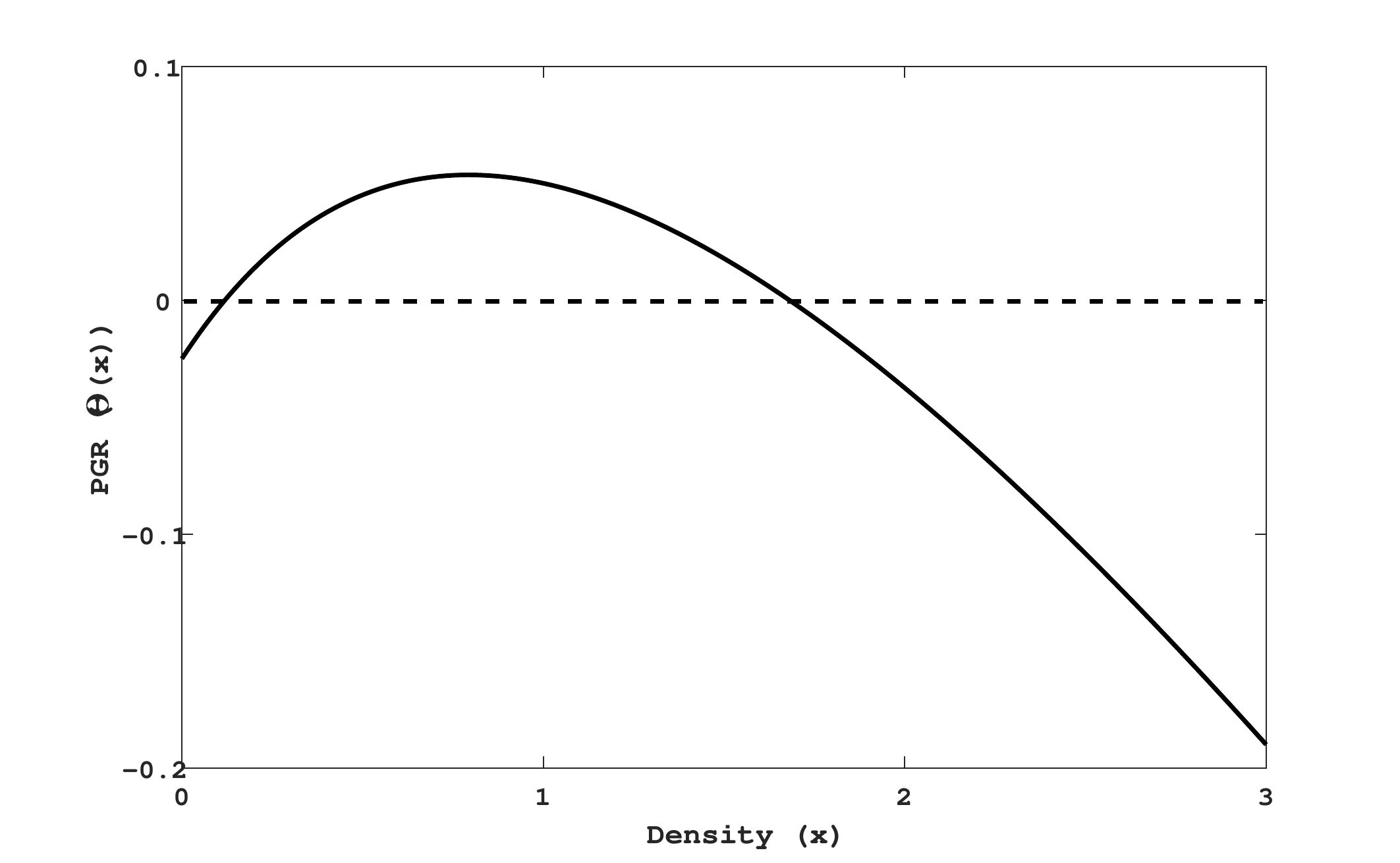}\label{fig_Strong_Allee}}
\subfigure[Model \eqref{model_single_species} shows logistic dynamics when $d_1(1+fy)(=1.3)<r(=2)<\frac{d_2(1+fy)^2}{cfy}(=4.69)$ for the parameter values $r=2$, $f=0.3$ and $c=0.3$.]{\includegraphics[height = 50mm, width = 70mm]{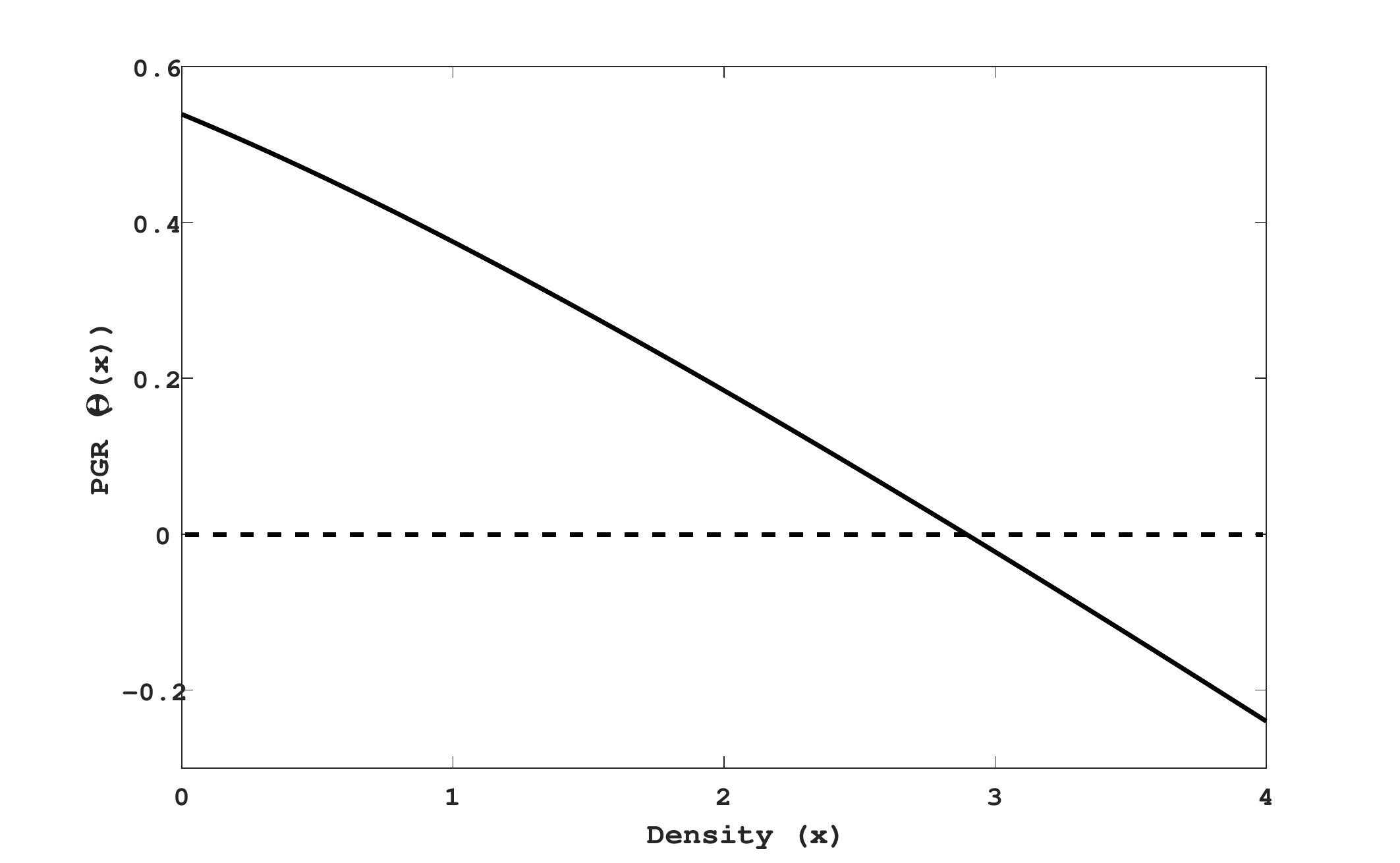}\label{fig_Logistic}}\hspace{10mm}
\subfigure[Model \eqref{model_single_species} shows logistic dynamics when either $f=0$ or $c=0$ (or both are zero) for the parameter values $r=2$, $f=0.3$ and $c=0.3$. Solid line is for $f=0$ (or when both are zero) and dot-dashed line is for $c=0$.]{\includegraphics[height = 50mm, width = 70mm]{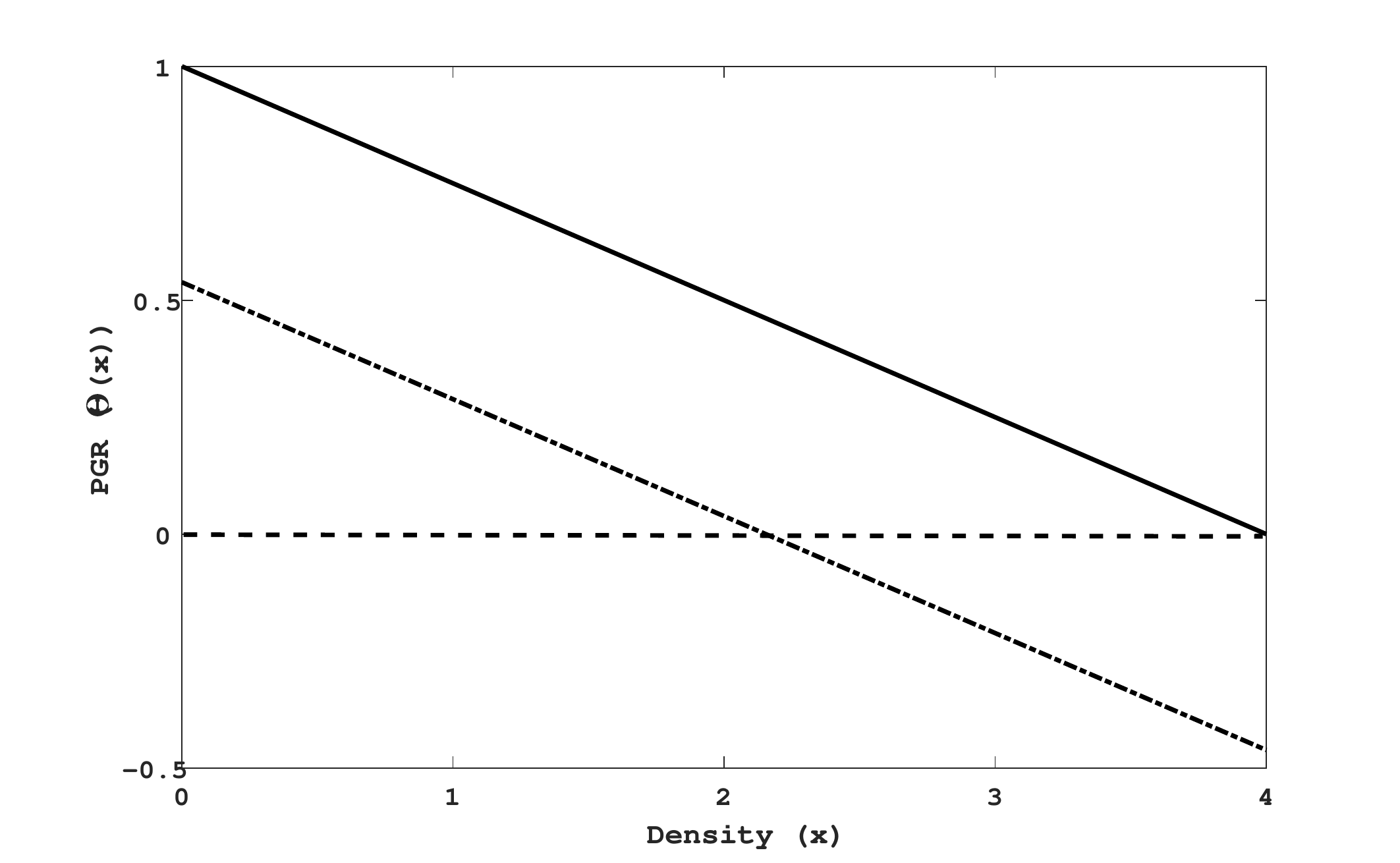}\label{fig_Logistic_f_c}}
\end{center}
\caption{Other parameters are fixed as $y=1$, $d_1=1$ and $d_2=0.25$. Dashed line represent the $x-axis$ (i.e., when PGR is zero).}
\label{fig_growth_dynamics_single_species}
\end{figure}

\begin{itemize}
\item If $r>d_1(1+fy)$, then the population of species is persistent in $\mathbb{R}_+$. Moreover, $\Theta(0)\geq0$ if $r\geq d_1(1+fy)$. Therefore, the Model \eqref{model_single_species} shows weak Allee dynamics (Figure \ref{fig_Weak_Allee}) if $$r>\max\Big\{d_1(1+fy), \frac{d_2(1+fy)^2}{cfy}\Big\}.$$

\item If $d_1+\frac{d_2(1+fy)}{c}<r<d_1(1+fy)$ and $\Delta>0$, then the population of species is persistent in $\mathbb{R}_+\backslash[0,x_2)$, and goes to extinction otherwise. Since $\Theta(0)<0$ if and only if $r<d_1(1+fy)$. The Model \eqref{model_single_species} shows strong Allee dynamics (Figure \ref{fig_Strong_Allee}) if the following conditions holds $$\max\Big\{d_1+\frac{d_2(1+fy)}{c},\frac{d_2(1+fy)^2}{cfy}\Big\}<r<d_1(1+fy) \mbox{ and } \Delta>0.$$ Here, $x_2$ is known as Allee threshold. Population will persist if initial population density is above $x_2$, otherwise it goes extinction. \\

\item Moreover, if $r<\frac{d_2(1+fy)^2}{cfy}$, then the System \eqref{model_single_species} shows logistic growth dynamics (Figure \ref{fig_Logistic}). Also, if $f=0$ or $c=0$ (or both are zero), then also equation \eqref{model_single_species} shows logistic growth dynamics (Figure \ref{fig_Logistic_f_c}).
\end{itemize}

\begin{rem}
In this section, we do not consider the predator dynamics explicitly. We only consider the constant predator population, without predation term. However, the qualitative properties of the Model \eqref{model_single_species}, will be the same for constant predator population model with Holling type $I$ functional response. In fact, we rewrite the model as,
\begin{eqnarray}\nonumber
\begin{array}{ccc}
\frac{dx}{dt} &=& rx\frac{1+cx}{1+cx+fy} - \underbrace{d_1}_{\text{constant death}}x - d_2x^2 - \underbrace{axy}_{\text{type-$I$ response}} \\
 &=& rx\frac{1+cx}{1+cx+fy} - \underbrace{(d_1+ay)}_{\text{constant death}}x - d_2x^2.
\end{array}
\end{eqnarray}
\end{rem}

In the next section, we will discuss the dynamics in the presence of predator population explicitly, with Holling type-$I$ functional response.

\section{Predator-prey model with type - I response function}\label{sec_predator-prey_type-I}
In this section, we study a prey-predator model with fear and carry-over effects with linear functional response (Holling type-$I$). Thus, our two-species prey-predator model becomes,
\begin{eqnarray}\label{model_two_species_type-1}
\begin{array}{ccc}
\frac{dx}{dt} &=& \frac{rx(1+cx)}{1+cx+fy} - d_1x - d_2x^2 - axy \\
\frac{dy}{dt} &=& a\alpha xy - my.
\end{array}
\end{eqnarray}
Here, $a$ is the rate of predation, $\alpha$ is the conversion efficiency from prey biomass to predator biomass and $m$ is the natural death rate of predator population. In the following theorem we summarize the basic dynamical properties of the Model \eqref{model_two_species_type-1}.

\begin{lemma}\label{thm_positivity_boundedness_two_species_type-I}[Positivity and boundedness of solutions for Model \eqref{model_two_species_type-1}.]
For the System \eqref{model_two_species_type-1}, the set $\mathbb{R}^2_+$ is positively invariant. Moreover, the System \eqref{model_two_species_type-1} is dissipative, i.e., every solution of \eqref{model_two_species_type-1} is ultimately bounded in $\mathbb{R}^2_+$, with the following properties
\begin{eqnarray}\nonumber
\begin{array}{ccc}
\lim_{t\rightarrow\infty}\sup x(t) \leq \frac{r-d_1}{d_2} \\\\
\lim_{t\rightarrow\infty}\sup \lz x(t)+\frac{1}{\alpha}y(t)\rz &\leq& \biggl\{^{\frac{r-d_1}{d_2} \mbox{~~~~~~~~~~ if } m>r-d_1}_{\frac{(r-d_1+m)^2}{4d_2m} \mbox{~~~~ if } m\leq r-d_1.}
\end{array}
\end{eqnarray}
\end{lemma}

The Model \eqref{model_two_species_type-1} always has the trivial extinction equilibrium $E_0 = (0,0)$. Apart from this, it has the boundary equilibrium (or prey only equilibrium) $E_1 = \lx\frac{r-d_1}{d_2},0\rx$, under the condition $r>d_1$. Moreover, there exists a unique interior equilibrium $E_* = \lx x_*,y_*\rx = \lx\frac{m}{a\alpha},y_*\rx$ where $y_*$ is the root of the quadratic equation $$\Gamma_0y^2+\Gamma_1y+\Gamma_2 = 0,$$ where
\begin{eqnarray}\nonumber
\begin{array}{ccc}
\Gamma_0 &=& a^3\alpha^2f \\
\Gamma_1 &=& a\alpha\lz a(a\alpha+cm)+f(d_1a\alpha+d_2m)\rz \\
\Gamma_2 &=& (a\alpha+cm)(d_1a\alpha+d_2m-ra\alpha).
\end{array}
\end{eqnarray}
The above equation has unique real positive root
\begin{eqnarray}\label{expression_equilibrium_density_y_type-I}
\begin{array}{ccc}
y_* = \frac{-\lz a(a\alpha+cm)+f(d_1a\alpha+d_2m)\rz+\sqrt{\lz a(a\alpha+cm)-f(d_1a\alpha+d_2m)\rz^2+4a^2\alpha rf(a\alpha+cm)}}{2a^2\alpha f},
\end{array}
\end{eqnarray}

\begin{eqnarray}\label{cond_existence_interior_type-I}
\begin{array}{ccc}
\mbox{ iff } ra\alpha>d_1a\alpha+d_2m \mbox{ i.e., iff } r>d_1+\frac{d_2m}{a\alpha}.
\end{array}
\end{eqnarray}

\begin{rem}
When the prey and predator population coexists at $E_*$, the density of prey population does not depend on any of the parameters, cost of predation fear $(f)$ and carry over effect $(c)$. However, the density of predator population depends on both the parameters. We have,
\begin{eqnarray}\label{eqn_equi_density_predator_partial_wrt_f}
\begin{array}{ccc}
\frac{\partial y_*}{\partial f} = -\frac{(a\alpha+cm)\lz a(a\alpha+cm)-f(d_1a\alpha+d_2m)+2a\alpha rf-\sqrt{\lz a(a\alpha+cm)-f(d_1a\alpha+d_2m)\rz^2+4a^2\alpha rf(a\alpha+cm)}\rz}{2a\alpha f^2\sqrt{\lz a(a\alpha+cm)-f(d_1a\alpha+d_2m)\rz^2+4a^2\alpha rf(a\alpha+cm)}}.
\end{array}
\end{eqnarray}
Now, it is easy to prove that the numerator of \eqref{eqn_equi_density_predator_partial_wrt_f} is positive, iff $r>d_1+\frac{d_2m}{a\alpha}$, which is the existence condition of $y_*$. Thus, $\frac{\partial y_*}{\partial f}<0$, i.e., at the coexistence state, as we increase the cost of fear, the density of predator population decreases. \\

Similarly, we have
\begin{eqnarray}\label{eqn_equi_density_predator_partial_wrt_c}
\begin{array}{ccc}
\frac{\partial y_*}{\partial c} = -\frac{m\lz\sqrt{\lz a(a\alpha+cm)-f(d_1a\alpha+d_2m)\rz^2+4a^2\alpha rf(a\alpha+cm)}-a(a\alpha+cm)+f(d_1a\alpha+d_2m)-2a\alpha rf\rz}{2a\sqrt{\lz a(a\alpha+cm)-f(d_1a\alpha+d_2m)\rz^2+4a^2\alpha rf(a\alpha+cm)}}.
\end{array}
\end{eqnarray}
Again, it is easy to prove that the numerator of \eqref{eqn_equi_density_predator_partial_wrt_c} is negative, iff $r>d_1+\frac{d_2m}{a\alpha}$, which is the existence condition of $y_*$. Thus, $\frac{\partial y_*}{\partial c}>0$, i.e., at the coexistence state, as we increase the carry over effect, the density of predator population increases.
\end{rem}

The following theorem describes the local stability of all three equilibria.

\begin{thm}\label{thm_local_stability_type-I}[Local stability of equilibria for Model \eqref{model_two_species_type-1}.]
\begin{enumerate}
\item The extinction equilibrium $E_0$ is locally asymptotically stable if $r<d_1$, and a saddle otherwise. \\

\item The prey-only equilibrium $E_1$ is locally asymptotically stable if the condition \eqref{cond_existence_interior_type-I} is reversed and a saddle when the condition \eqref{cond_existence_interior_type-I} is satisfied. \\

\item The coexistence equilibrium $E_*$ is locally asymptotically stable if $r<\frac{d_2(a\alpha+cm+a\alpha fy_*)^2}{a^2\alpha^2cfy_*}$, where $y_*$ is defined in \eqref{expression_equilibrium_density_y_type-I}.\\

Moreover, if $d_1 + \frac{d_2m}{a\alpha} < r < \frac{4d_2(a\alpha+cm)}{a\alpha c}$, then the coexistence equilibrium $E_*$ is always locally asymptotically stable.
\end{enumerate}
\end{thm}

The following theorems give the additional stability properties for the coexistence equilibrium.

\begin{thm}\label{thm_critical_value_of_f}
If $r<r_c$, then the coexistence equilibrium $E_*$ is always locally asymptotically stable. Furthermore, if $r>r_c$, then the sufficient condition for the local stability of the coexistence equilibrium $E_*$ is $f<f_c$, where $r_c$ and $f_c$ are given by
\begin{eqnarray}\nonumber
\begin{array}{ccc}
r_{c} &=& \frac{c(a\alpha+cm)(d_1a\alpha+d_2m)+\sqrt{c(a\alpha+cm)(d_1a\alpha+d_2m)\lz c(a\alpha+cm)(d_1a\alpha+d_2m)+4a^2\alpha^2d_2\rz}}{2a\alpha c(a\alpha+cm)}, \\
f_c &=& \frac{a^2\alpha d_2\lz a\alpha d_2+rc(a\alpha+cm)\rz}{rc\lz a\alpha r^2c(a\alpha+cm)-rc(a\alpha+cm)(d_1a\alpha+d_2m)-a\alpha d_2(d_1a\alpha+d_2m)\rz}.
\end{array}
\end{eqnarray}
\end{thm}

\begin{rem}
If $r<r_c$, then the condition \eqref{cond_sufficient_stability_interior_type-I} is always satisfied, i.e., when $r$ is small (smaller than $r_c$), the fear parameter $f$ has no role in the coexistence equilibrium stability. In other words, if $$d_1+\frac{d_2m}{a\alpha}<r<r_c,$$ then the equilibrium $E_*$ is always locally asymptotically stable. The stability of $E_*$ will not change if the birth rate of prey is not large enough to support oscillations, which is similar to the result obtained by \citet{wang2016modelling} with type-$II$ response function. \\

If $r>r_c$, then the condition \eqref{cond_sufficient_stability_interior_type-I} is satisfied if $f<f_c$, which is the sufficient condition for the stability of $E_*$, when $r$ is large (larger than $r_c$). In other words, if $$r > \max\Big\{d_1+\frac{d_2m}{a\alpha}, r_c\Big\},$$ then the sufficient condition for the local stability of $E_*$ is $f<f_c$. \\

Therefore, we can see that the fear parameter has an important role on the stability of the coexistence equilibrium, even for the type-$I$ response function. However, the actual stability property with respect to the fear parameter $f$ is more complex than the above scenarios, which is discussed later through numerical simulations.
\end{rem}

\begin{thm}\label{thm_critical_value_of_C}
If $r<\frac{4d_2m}{a\alpha}$, then the coexistence equilibrium $E_*$ is locally asymptotically stable. Furthermore, if $r>\frac{4d_2m}{a\alpha}$, then the sufficient condition for the local stability of $E_*$ is $c<c_r$, which does not depend on the fear parameter $f$, where $\lx c_r = \frac{4d_2a\alpha}{ra\alpha-4d_2m}\rx$.
\end{thm}

\begin{rem}
If $r<\frac{4d_2m}{a\alpha}$, then the condition \eqref{cond_sufficient_stability_interior_type-I_for_c} is always satisfied, i.e., when $r$ is small, the parameter $c$ has no role in the stability of the coexistence equilibrium. In other words, if $$d_1+\frac{d_2m}{a\alpha}<r<\frac{4d_2m}{a\alpha},$$ then the equilibrium $E_*$ is always locally asymptotically stable. The stability of $E_*$ will not be changed if the birth rate of prey is not large enough to support oscillations. \\

If $r>\frac{4d_2m}{a\alpha}$, then the condition \eqref{cond_sufficient_stability_interior_type-I_for_c} is satisfied if $c<c_r$, which is the sufficient condition for the stability of $E_*$. In other words, if $$r>\max\Big\{d_1+\frac{d_2m}{a\alpha},\frac{4d_2m}{a\alpha}\Big\},$$ then the sufficient condition for the local stability of $E_*$ is $c<c_r$ \eqref{cond_c_crit}.
\end{rem}

From the previous two Theorems \eqref{thm_critical_value_of_f} and \eqref{thm_critical_value_of_C}, we can see that when prey growth rate $r$ is small, the stability of the coexistence equilibrium is not affected by the cost of fear or carry over effect. For the classical prey-predator model with Holling type $I$ functional response, without cost of fear and carry over effect, Hopf-bifurcation never occur (our Model \eqref{model_two_species_type-1} will be reduced to the classical prey-predator model with Holling type $I$ functional response, if we neglect the cost of fear). The result is same for the prey-predator model with Holling type $I$ functional response with only the cost of fear. However, for our model with both cost of fear and carry over effect, Hopf-bifurcation occurs as we increase the parameter $r$, and the phenomenon \textit{'paradox of enrichment'} appears \citep{mcallister1972stability, riebesell1974paradox, rosenzweig1971paradox, gilpin1972enriched}. When birth rate of prey is large enough, prey and predator can still go to the coexistence steady state according to preys cost of fear or carry over effect. If either of the cost of fear and carry over effect is small enough, then it can suppress oscillations. Therefore, by incorporating the cost of fear and carryover effect, the phenomenon \textit{'paradox of enrichment'} can occur, however, we can rule out such phenomenon by choosing suitable $f$ or $c$. It is to be noted that the carrying capacity (e.g. for logistic model) is considered as the parameter to be evaluated in terms of enrichment \citep{morozov2007towards}. In this study, the parameter $r$, which is the maximum birth rate of prey, is considered as the parameter to be evaluated in terms of enrichment. Actually, if we simplify the Model \eqref{model_logistic} then we will get the expression for the carrying capacity as $\frac{r-d_1}{d_2}$. From this expression, we can see that the carrying capacity increases or decreases with the parameter $r$, when other parameters are fixed. Therefore, it is reasonable to assume the parameter $r$ as the parameter to be evaluated in terms of enrichment. \\

All the local stability conditions in the Theorem \eqref{thm_local_stability_type-I} are actually global conditions. In the next theorems we will discuss about the global stability of all equilibria for Model \eqref{model_two_species_type-1}.

\begin{thm}\label{thm_global_stability_boundary_equi_type-I}[Global stability of boundary equilibria for Model \eqref{model_two_species_type-1}.]
The equilibrium $E_0$ is globally asymptotically stable if $r\in(0,d_1)$ and $E_1$ is globally asymptotically stable if $r\in\lx d_1,d_1+\frac{d_2m}{a\alpha}\rx$.
\end{thm}

\begin{thm}\label{thm_global_stability_interior_equi_type-I}[Global stability of interior equilibrium for Model \eqref{model_two_species_type-1}.]
The positive equilibrium $E_*$ is globally asymptotically stable if $r\in\lx d_1+\frac{d_2m}{a\alpha}, \frac{4d_2}{c}\rx$.
\end{thm}

\begin{rem}
From the Theorem \eqref{thm_local_stability_type-I}, we can see that, as the parameter $r$ increases the Model \eqref{model_two_species_type-1} experiences two bifurcation of equilibrium and a Hopf-bifurcation at positive equilibrium (discussed later). When $0<r<d_1$, $E_0$ is globally asymptotically stable, when $r$ passes $d_1$, $E_0$ loses its stability to $E_1$, which becomes globally asymptotically stable in $d_1<r<d_1+\frac{d_2m}{a\alpha}$. Again, when $r$ passes $d_1+\frac{d_2m}{a\alpha}$ then $E_1$ loses its stability to $E_*$, which is locally asymptotically stable in $d_1+\frac{d_2m}{a\alpha}<r<\frac{d_2(a\alpha+cm+a\alpha fy_*)^2}{a^2\alpha^2cfy_*}$, where $y_*$ is defined in \eqref{expression_equilibrium_density_y_type-I}. Moreover, when $r$ passes through $\frac{d_2(a\alpha+cm+a\alpha fy_*)^2}{a^2\alpha^2cfy_*}$, then $E_*$ loses its stability and limit cycle oscillation occurs around $E_*$ through Hopf-bifurcation.
\end{rem}

\subsection{Existence of limit cycles and Hopf-bifurcation}
In this subsection, we investigate the possibility of Hopf-bifurcation at the coexistence equilibrium $E_*$ by considering the fear effect parameter $f$, as the bifurcation parameter. Similarly, we can obtain the Hopf-bifurcation with respect to other parameters $r$ and $c$. \\

At the Hopf-bifurcation point, the real parts of the eigenvalues of the characteristic equation \eqref{eqn_char_int_type-I} equal to zero. We set, at $f=f_H$ the Hopf-bifurcation occurs, which gives $$\Psi_{11}(f_H) = 0 \mbox{ and } \Psi_{12}(f_H)\Psi_{21}(f_H)>0.$$ Thus, at the Hopf-bifurcation point, we have $$\Psi_{11}(f_H) = 0 \Rightarrow f_H = \frac{a\alpha rc-2d_2(a\alpha+cm)\pm\sqrt{a\alpha rc\lz a\alpha rc-4d_2(a\alpha+cm)\rz}}{2d_2a\alpha y_*}.$$

Moreover, if we simplify the above condition by using maple software, we can obtain that the values of $f_H$ are the roots of the quadratic equation $$C_1f^2 + C_2f + C_3 = 0,$$ where

\begin{eqnarray}\nonumber
\begin{array}{ccc}
C_1 &=& d_2\lz c(d_1a\alpha+d_2m)(ar\alpha-(d_1a\alpha+d_2m)) - ar\alpha d_2(a\alpha+cm)\rz \\
C_2 &=& ac\lz arc\alpha(ar\alpha-(d_1a\alpha+d_2m)) + d_2(a\alpha+cm)(2(d_1a\alpha+d_2m)-3ar\alpha)\rz \\
C_3 &=& -a^2cd_2(a\alpha+cm)^2.
\end{array}
\end{eqnarray}

Therefore, $f_H = \frac{-C_2\pm\sqrt{C_2^2-4C_1C_3}}{2C_1}.$ \\

Further, at the Hopf-bifurcation point, we have $$\frac{d(\lambda_r)}{df}\Big|_{f=f_H} = \frac{2\lambda_i^2\frac{d(\Psi_{11})}{df}-\Psi_{11}\frac{d(\Psi_{12}\Psi_{21})}{df}}{\Psi_{11}^2+4\lambda_i^2} \Big|_{f=f_H}\neq0,$$ which is true if $\lz2\lambda_i^2\frac{d(\Psi_{11})}{df}-\Psi_{11}\frac{d(\Psi_{12}\Psi_{21})}{df}\rz\Big|_{f=f_H}\neq0$, where $\lambda_r$ and $\lambda_i$ are real and imaginary parts of the eigenvalues of the characteristic equation \eqref{eqn_char_int_type-I}. \\

The following theorem gives conditions for the existence of Hopf-bifurcation at $E_*$ for Model \eqref{model_two_species_type-1}. \\

\begin{thm}\label{thm_cond_existence_Hopf-bifurcation}[Condition for the existence of Hopf-bifurcation at interior equilibrium for Model \eqref{model_two_species_type-1}.]
If $\Psi_{12}(f_H)\Psi_{21}(f_H)>0$, and $\lz2\lambda_i^2\frac{d(\Psi_{11})}{df}-\Psi_{11}\frac{d(\Psi_{12}\Psi_{21})}{df}\rz\Big|_{f=f_H}\neq0$, hold, then the interior equilibrium $E_*$ of Model \eqref{model_two_species_type-1} is locally asymptotically stable when $f<f_H$, and undergoes Hopf-bifurcation at $E_*$ when $f=f_H$.
\end{thm}

The following theorem gives the direction and stability of Hopf-bifurcation around the coexistence steady state $E_*$.

\begin{thm}\label{thm_cond_direction-stability_Hopf-bifurcation}[Direction and stability of Hopf-bifurcation at interior equilibrium for Model \eqref{model_two_species_type-1}.]
Let define $L$ as
\begin{eqnarray}\nonumber
\begin{array}{ccc}
L &:=& 3g_v^2g_{uuu}\lz(1-g_{uv}z_1)+2z_1(g_{uvv}z_2w-g_{uuv}z_1g_v)\rz + 3wg_{vvv}z_2\lz2w(h_ug_{uvv}z_2-wg_{uuv}z_1)+g_{uv}h_u\rz \\
&+& g_{uvv}w\lz w(1-g_{uv}z_1)+2z_2(g_vg_{uu}-h_ug_{vv})+2z_2w(wg_{uvv}z_1-g_vz_2g_{uuv})\rz \\
&+& 2g_{uuv}z_1\lz g_v^2(wg_{uuv}z_2-g_{uu})-w^2(g_{vv}+g_vz_1g_{uvv})\rz + g_{uv}(h_ug_{vv}+wg_vg_{uuv}z_2-g_vg_{uu}).
\end{array}
\end{eqnarray}

Then the Hopf-bifurcation is supercritical if $L<0$ and it is subcritical if $L>0$.
\end{thm}

\section{Numerical simulations}\label{sec_numerical}

\begin{figure}
\begin{center}
\subfigure[Bifurcation diagram of prey w.r.t. $r$.]{\includegraphics[height = 50mm, width = 70mm]{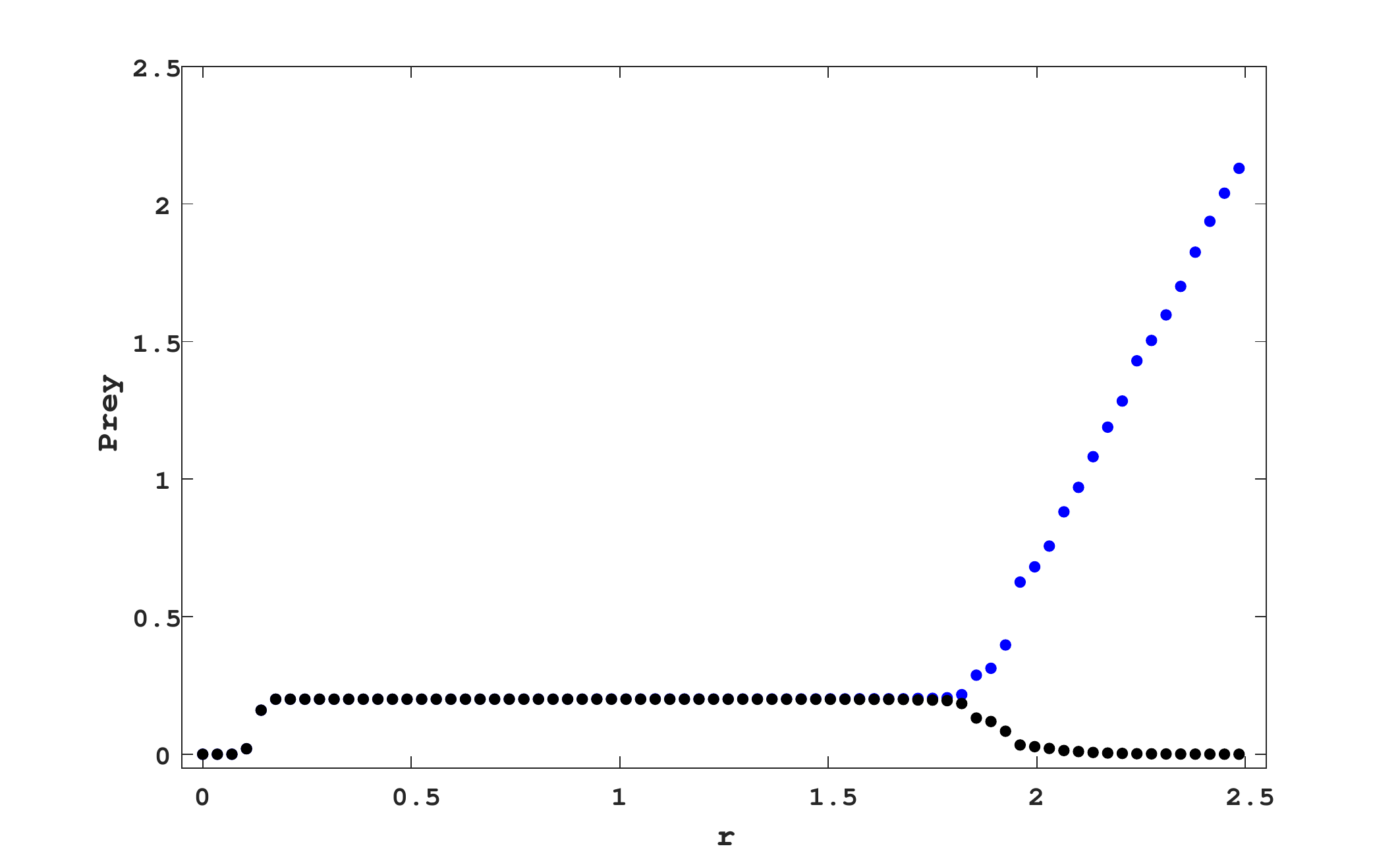}\label{fig_r_vs_prey}}\hspace{10mm}
\subfigure[Bifurcation diagram of predator w.r.t. $r$.]{\includegraphics[height = 50mm, width = 70mm]{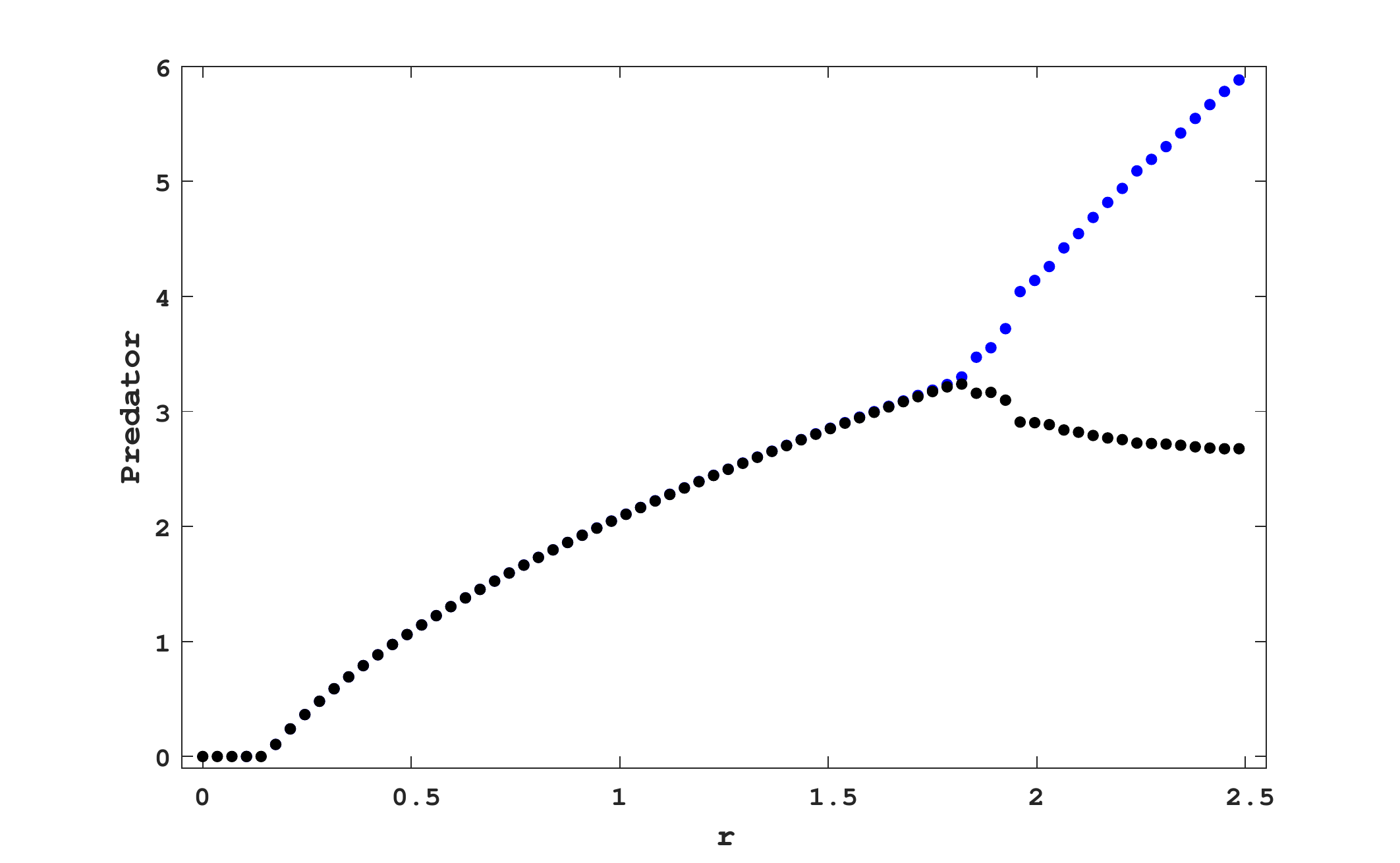}\label{fig_r_vs_predator}}
\end{center}
\caption{Bifurcation diagrams for Model \eqref{model_two_species_type-1} with respect to $r$. Here, $r$ varies from $0$ to $2.5$, with the initial condition $[x(0), y(0)] = [0.5, 0.1]$. Fixed parameter values are $f = 1$, $c = 0.8$, $d_1 = 0.1$, $d_2 = 0.25$, $a = 0.1$, $\alpha = 0.5$, and $m = 0.01$.}
\label{fig_bifurcation_wrt_r_type-I}
\end{figure}

In the Figure \eqref{fig_bifurcation_wrt_r_type-I}, we fix the parameters $f$, $c$, $d_1$, $d_2$, $a$ , $\alpha$ and $m$ (the specific values are given in the figure) and vary the parameter $r$. From the figure, we can see that as we increase the parameter $r$, the coexistence equilibrium becomes unstable through Hopf-bifurcation and prey-predator oscillation occurs. This phenomenon is known as \textit{'paradox of enrichment},' which can not be observed in the classical prey-predator model with Holling type $I$ functional response (even in the presence of only fear effect \citep{wang2016modelling, sasmal2018population}). Here, we consider the parameter $r$ as the enrichment parameter because, if we simplify the Model \eqref{model_logistic}, we will get the expression for carrying capacity as $\frac{r-d_1}{d_2}$, therefore, it is reasonable to assume $r$ as the enrichment parameter, when other parameters are fixed. \\

\begin{figure}
\begin{center}
\subfigure[Bifurcation diagram of prey w.r.t. $f$.]{\includegraphics[height = 50mm, width = 70mm]{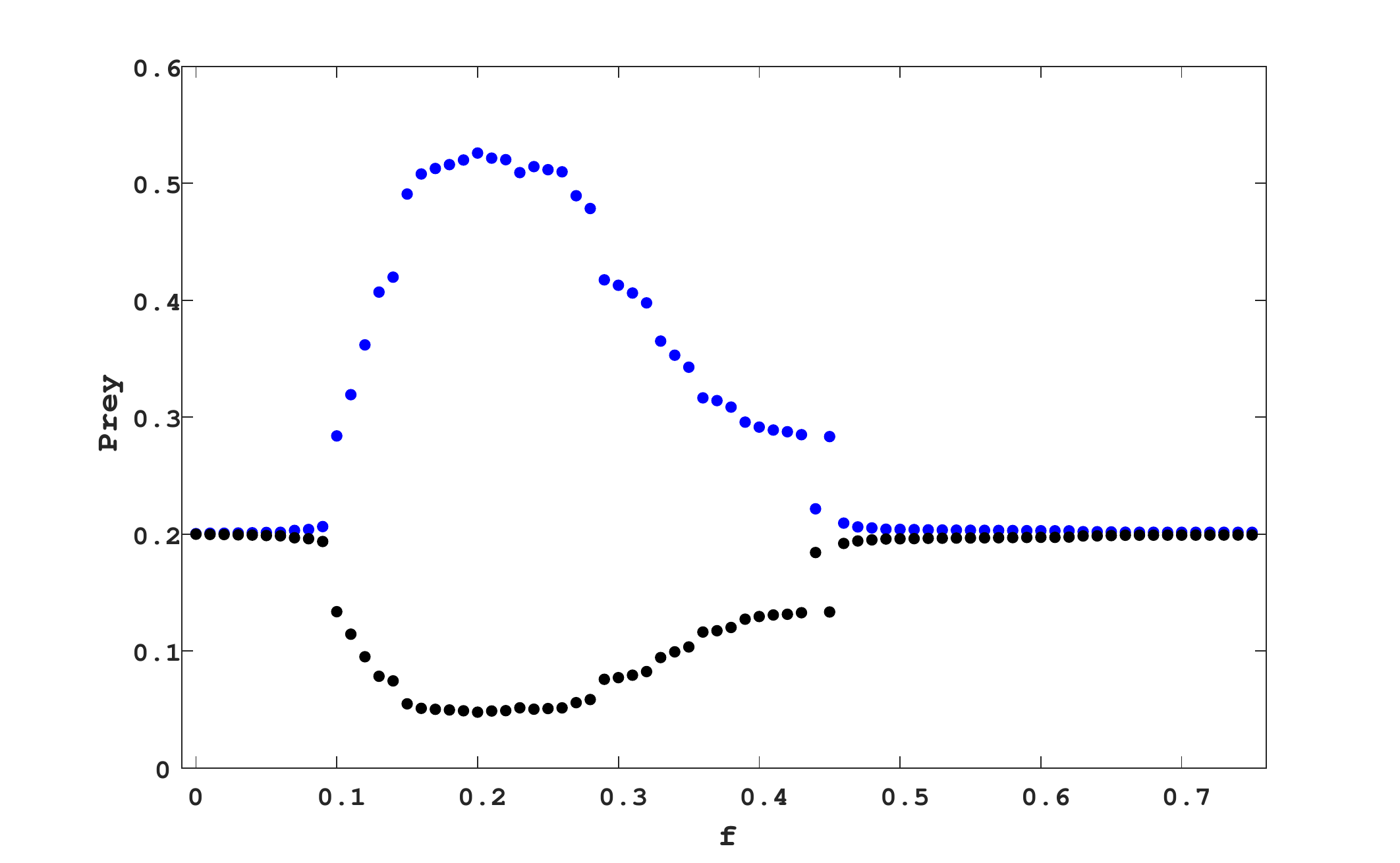}\label{fig_f_vs_Prey}}\hspace{10mm}
\subfigure[Bifurcation diagram of predator w.r.t. $f$.]{\includegraphics[height = 50mm, width = 70mm]{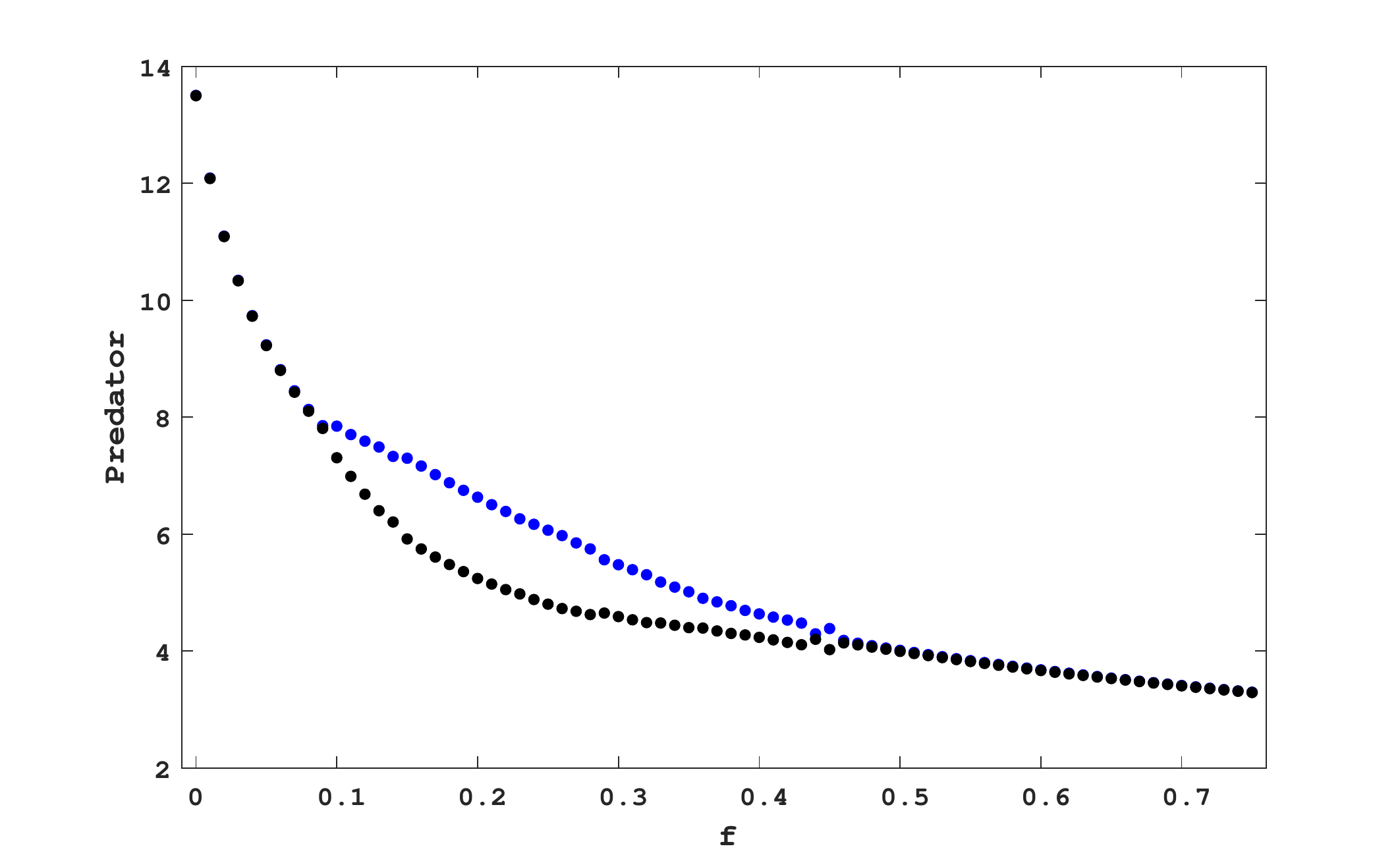}\label{fig_f_vs_Predator}}
\end{center}
\caption{Bifurcation diagrams for Model \eqref{model_two_species_type-1} with respect to $f$. Here, $f$ varies from $0$ to $0.75$, with the initial condition $[x(0), y(0)] = [0.5, 0.1]$. Fixed parameter values are $r = 1.5$, $c = 0.8$, $d_1 = 0.1$, $d_2 = 0.25$, $a = 0.1$, $\alpha = 0.5$, and $m = 0.01$. The critical values of $r$ and $f$ are $r_c = 0.2896$ \eqref{cond_r_crit} and $f_c = 0.0186$ \eqref{cond_f_crit}, respectively.}
\label{fig_bifurcation_wrt_f_type-I}
\end{figure}

\begin{figure}
\begin{center}
\subfigure[Bifurcation diagram of prey w.r.t. $c$.]{\includegraphics[height = 50mm, width = 70mm]{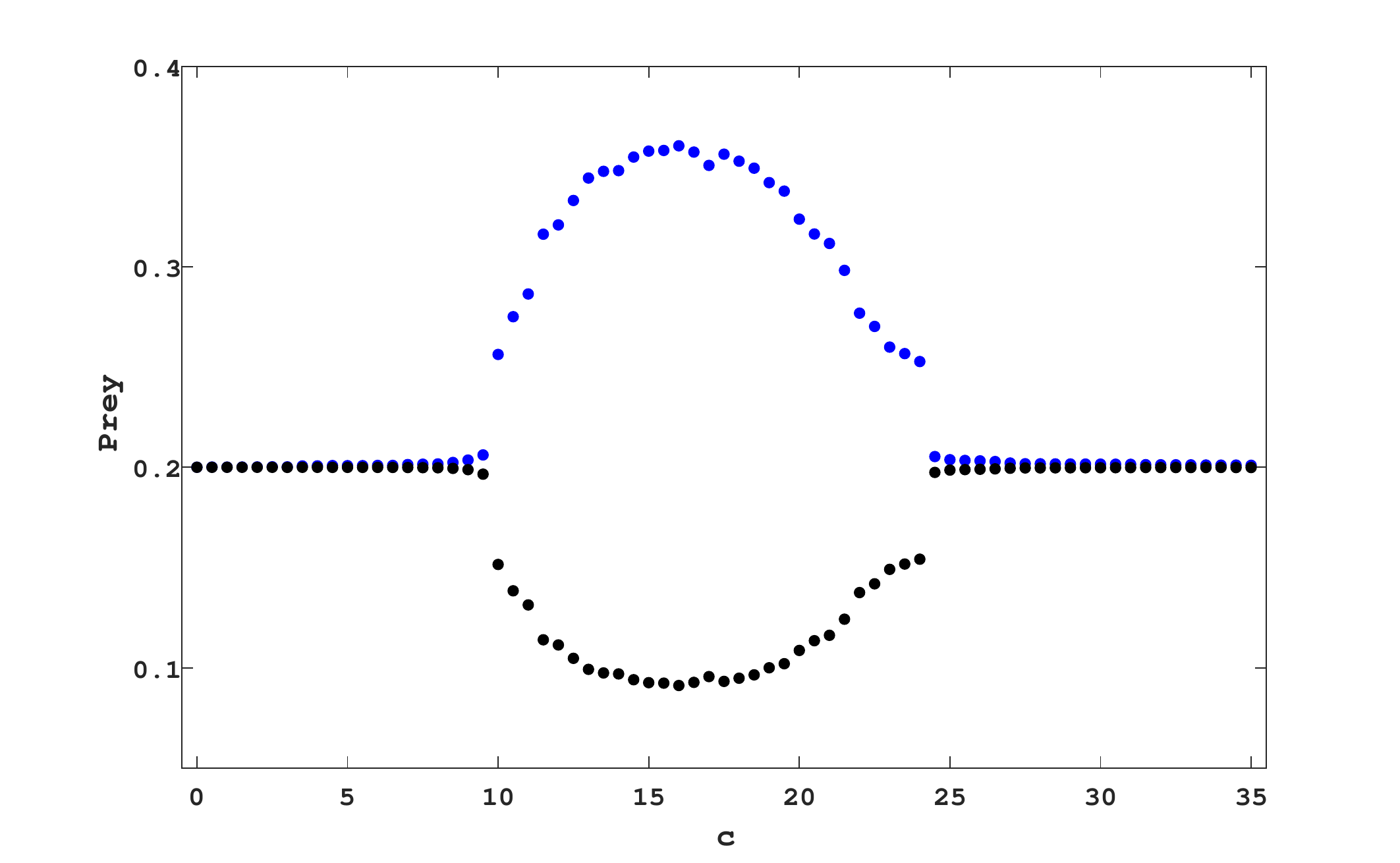}\label{fig_c_vs_prey}}\hspace{10mm}
\subfigure[Bifurcation diagram of predator w.r.t. $c$.]{\includegraphics[height = 50mm, width = 70mm]{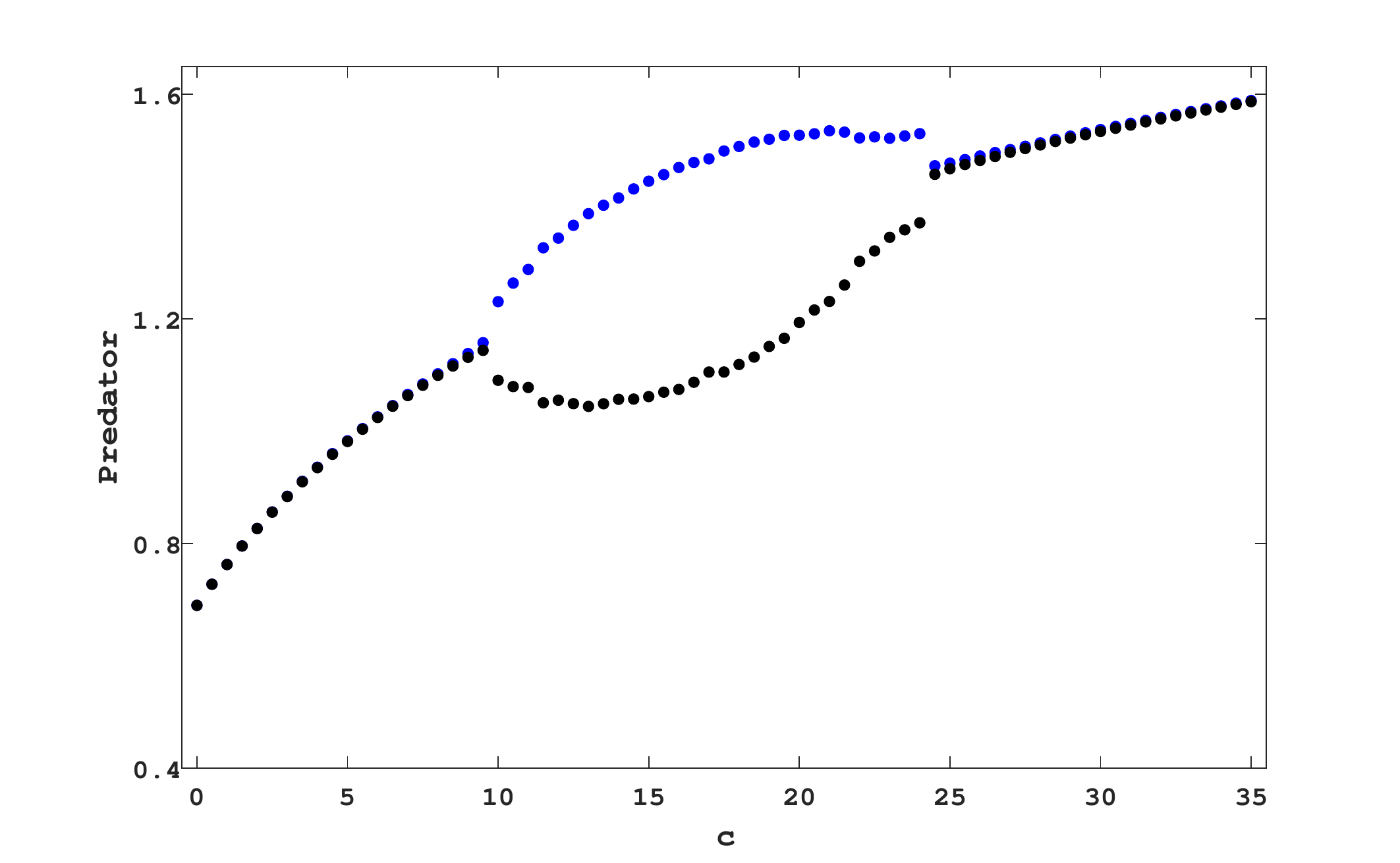}\label{fig_c_vs_predator}}
\end{center}
\caption{Bifurcation diagrams for Model \eqref{model_two_species_type-1} with respect to $c$. Here, $c$ varies from $0$ to $35$, with the initial condition $[x(0), y(0)] = [0.5, 0.1]$. Fixed parameter values are $r = 0.37$, $f = 1$, $d_1 = 0.1$, $d_2 = 0.25$, $a = 0.1$, $\alpha = 0.5$, and $m = 0.01$. The critical values of $r = \frac{4d_2m}{a\alpha} = 0.2$ and $c = c_r = 5.8824$ \eqref{cond_c_crit}.}
\label{fig_bifurcation_wrt_c_type-I}
\end{figure}

\begin{figure}
\begin{center}
\subfigure[Stable limit cycle oscillation occur around the coexistence equilibrium $E_*$, for $f=1$.]{\includegraphics[height = 50mm, width = 70mm]{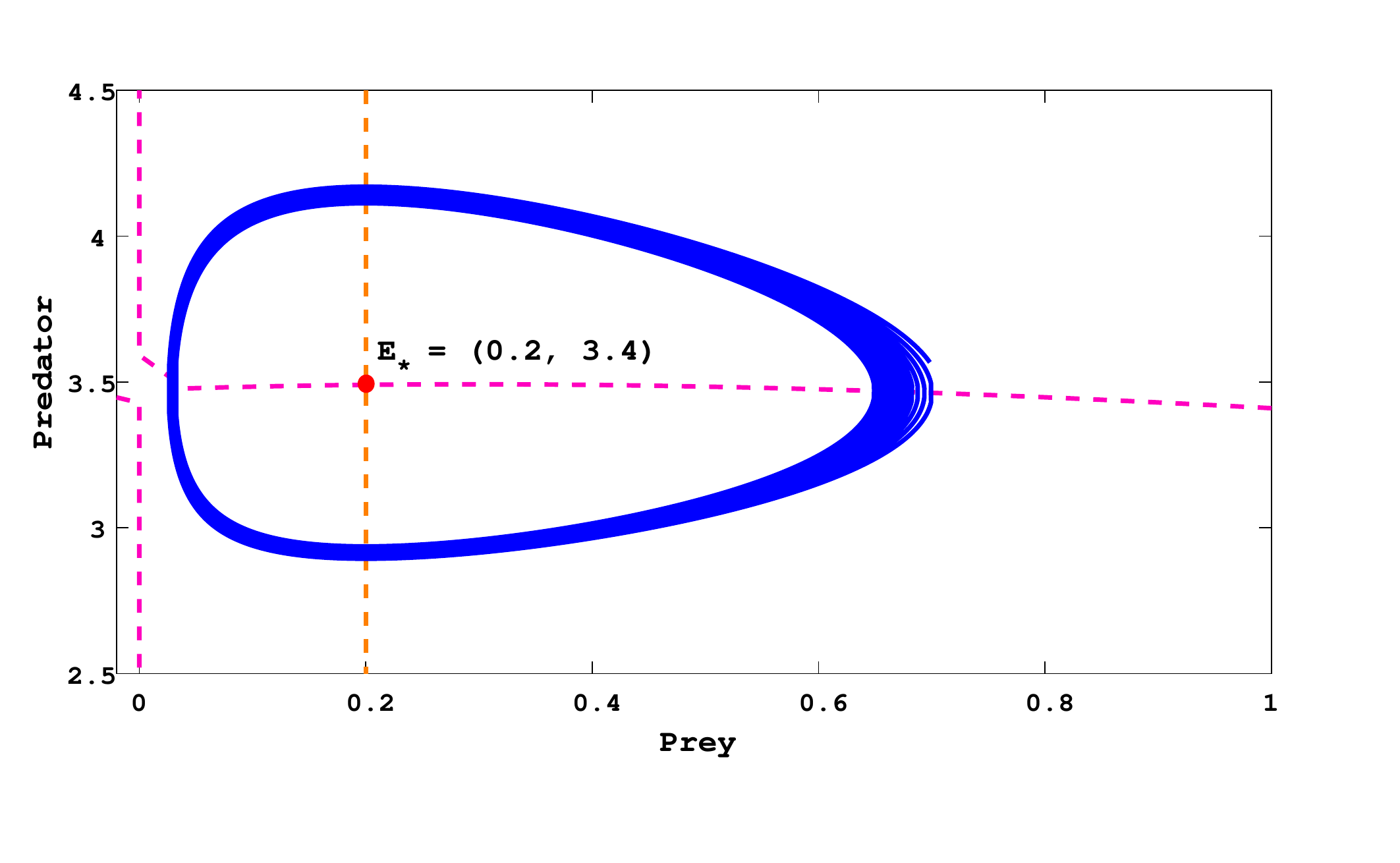}\label{fig_limit_cycle_for_r2f1}}\hspace{10mm}
\subfigure[Coexistence equilibrium is stable for $f=1.5$.]{\includegraphics[height = 50mm, width = 70mm]{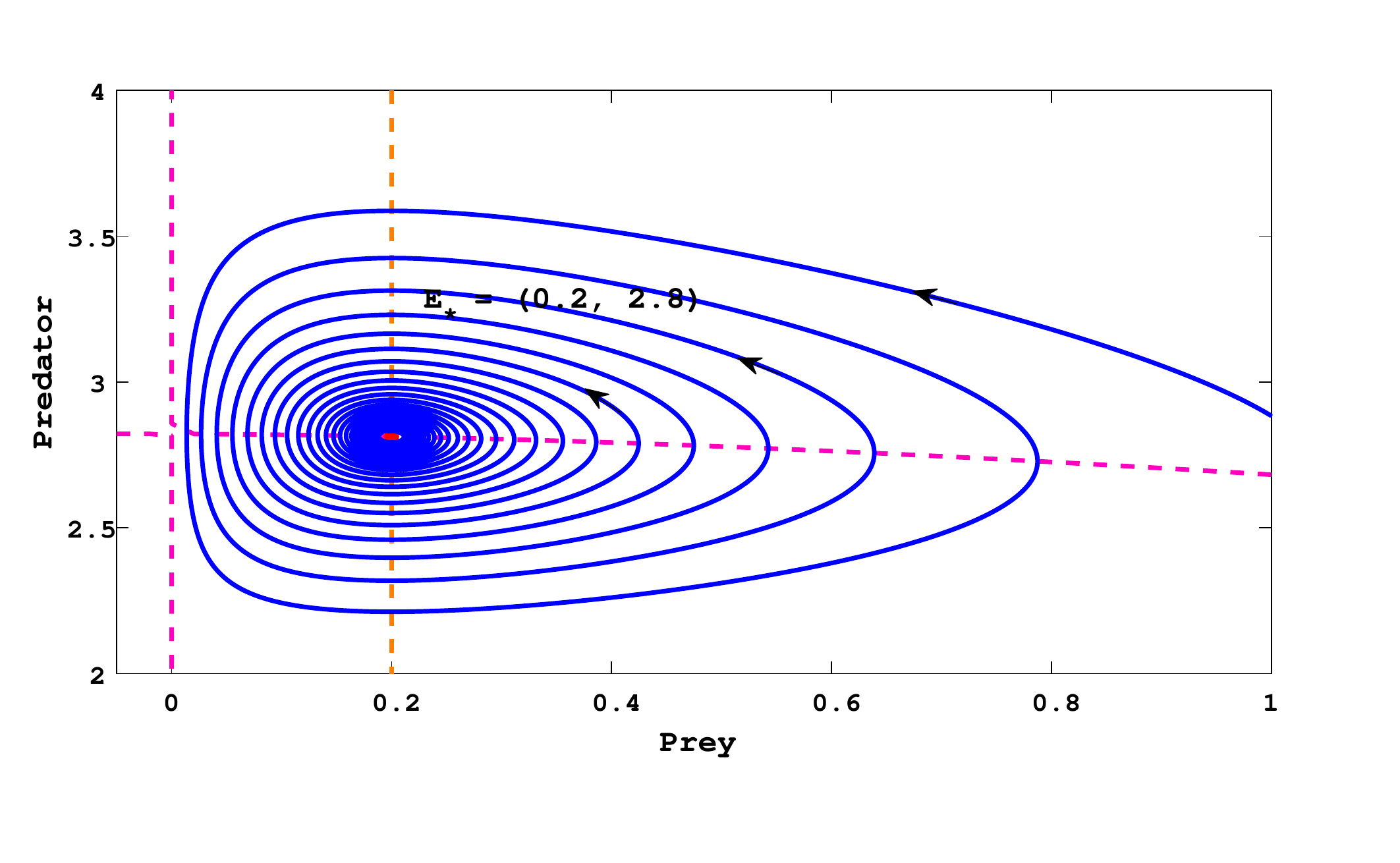}\label{fig_stable_interior_for_r2f1.5}}
\end{center}
\caption{Coexistence equilibrium stability for the fixed parameter values $r = 2$, $c = 0.8$, $d_1 = 0.1$, $d_2 = 0.25$, $a = 0.1$, $\alpha = 0.5$, and $m = 0.01$, corresponding to the Model \eqref{model_two_species_type-1}.}
\label{fig_behavior_type-I_wrt_r_and_f}
\end{figure}

\begin{figure}
\begin{center}
\includegraphics[height = 70mm, width = 120mm]{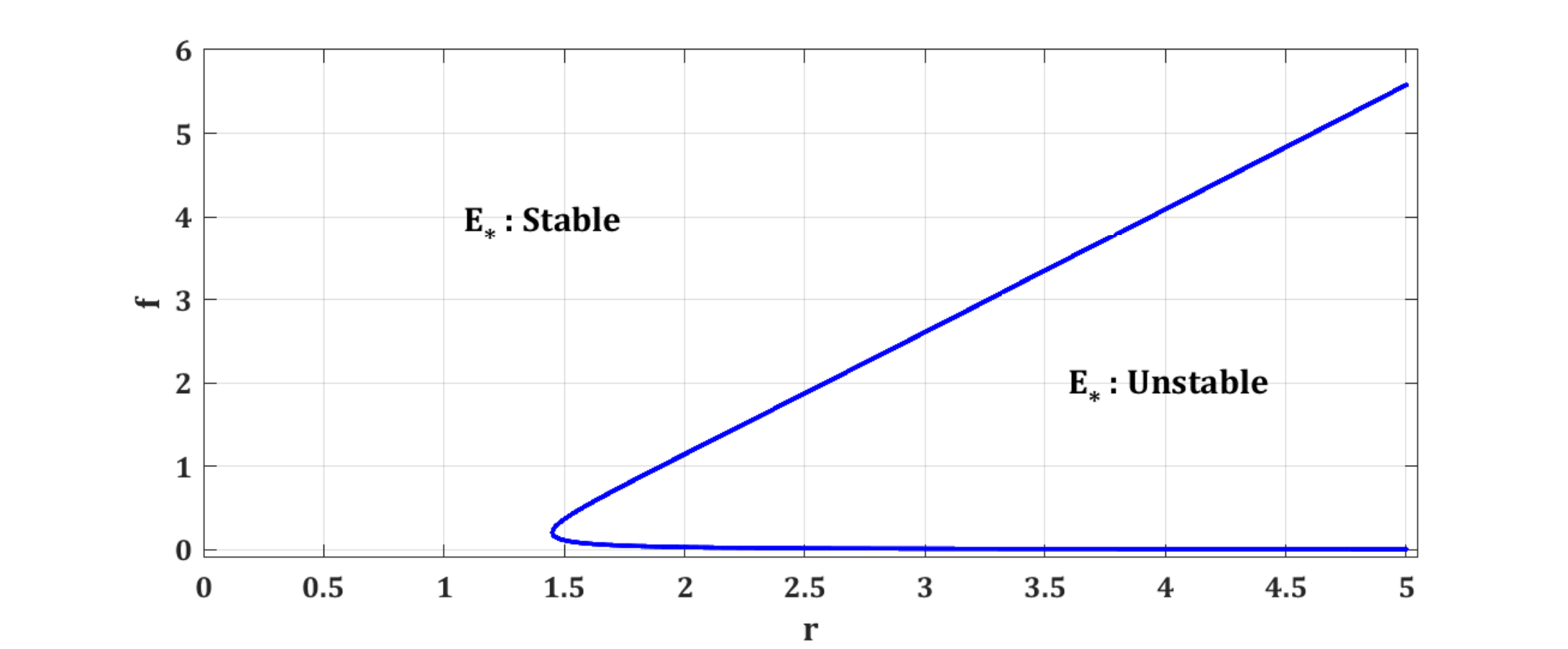}
\end{center}
\caption{Hopf-bifurcation curve in $r-f$ parameter plane for the Model \eqref{model_two_species_type-1}. Other parameter values are fixed at $c = 0.8$, $d_1 = 0.1$, $d_2 = 0.25$, $a = 0.1$, $\alpha = 0.5$, and $m = 0.01$.}
\label{fig_Hopf-Bif_curve_wrt_r&f}
\end{figure}

In Theorems \eqref{thm_critical_value_of_f} and \eqref{thm_critical_value_of_C}, we found the critical values of $r$ and have shown that if $r$ is less than the critical values then oscillation cannot be observed for our model \eqref{model_two_species_type-1}. The coexistence equilibrium is always stable irrespective of the non-lethal parameters values. For oscillation, $r$ must be greater than some threshold value, and in that case also, oscillation can be suppressed by the non-lethal effects parameters, which is shown in Figures \eqref{fig_bifurcation_wrt_f_type-I} and \eqref{fig_bifurcation_wrt_c_type-I}. Theoretically, we prove that for large values of $r$, if $f$ or $c$ is sufficiently small, then we can suppress the oscillation, however, the situation is more complex, which is shown in these two figures. In both the Figures \eqref{fig_bifurcation_wrt_f_type-I} and \eqref{fig_bifurcation_wrt_c_type-I}, we choose $r$ sufficiently large (greater than the critical values) and observed that predator-prey oscillation can occur only for the intermediate values of $f$ and $c$. Therefore, oscillation can be suppressed by both sufficiently low and high values of the cost of fear and carry over effect parameters, and we can rule out the phenomenon \textit{'paradox of enrichment.'} We draw phase diagrams in Figure \eqref{fig_behavior_type-I_wrt_r_and_f}, to show how we can rule out the oscillating behavior by choosing high anti-predator response. Moreover, in Figure \eqref{fig_Hopf-Bif_curve_wrt_r&f}, we have shown the Hopf-bifurcation curve for our Model \eqref{model_two_species_type-1}, in the $r-f$ parameter plane. From this figure, we can see that the parameter $r$ should be sufficiently large to show oscillatory behavior. Moreover, numerically we have checked the direction and stability of Hopf-bifurcation and found that the Hopf-bifurcation is only supercritical. We use Matlab 2017b software to produce all the figures (\eqref{fig_bifurcation_wrt_r_type-I} to \eqref{fig_Hopf-Bif_curve_wrt_r&f}) in this section.

\section{Discussion}\label{sec_discussion}
One of the central topics in ecology and evolutionary biology is to understand the variety of mechanisms which influence the fitness and survivability of the population \citep{pough1989organismal}. In the literature, most of the studies only concentrated on the lethal effects of predator in prey-predator interaction. However, some recent studies showed that apart from the lethal effects, there are some non-lethal effects, whose impact is equally important as of the previous one \citep{preisser2008many}. Among such effects, fear of predation has an important role on population fitness and survivability in the prey-predator system. Even such non-lethal effects are not restricted to affect in a single generation or in a particular season, it can be carried-over over generations or within generation \citep{o2014biological}. Therefore, in the present study, we consider both the cost of predation fear and its carry-over effects in population model. \\

First, we developed and analyzed a single species (e.g., prey) population model, by incorporating non-lethal effects of predator by considering constant predator population and without explicit predator dynamics. We incorporate the non-lethal effects in form of birth rate reduction due to the fear of predation. Moreover, we consider that such non-lethal effects can be carried within or over generations. We provide detailed analysis of our single species model, by neglecting direct predation, however, the qualitative properties of the model will be the same, if we consider predation followed by the Holling type $I$ functional response. We derive the global dynamical properties of our proposed single species Model \eqref{model_single_species}. For the single species Model \eqref{model_single_species}, our main study objective is to find the different growth dynamics due to the cost of predation fear and its carry-over effects. More specifically, our goal is to investigate the occurrence of Allee effects due to such non-lethal effects of predator. From our analysis, we can see that such non-lethal effects can be a cause of generating Allee effects and our model shows three types of growth dynamics; namely, weak Allee dynamics, strong Allee dynamics and logistic dynamics, depending on the restrictions of model parameters, which is summarized in the Table \eqref{table_growth_dynamics_single-species-model}. The above results not only relate the Allee effects and predation fear mechanism, but also answer the first objective raised in the introduction section. Moreover, our system shows both saddle-node and transcritical bifurcations depending on the parameter values. \\

Next, we include the explicit dynamics of predator population, where predation follows Holling type $I$ \citep{Holling1959} functional response. We derive the basic dynamical properties of our proposed model, existence and local stability conditions of each equilibria. Unlike the previous studies \citep{wang2016modelling, sasmal2018population}, our mathematical and numerical results show that the cost of predation fear and its carry-over effect, affect the prey-predator interactions in many ways, even if predation is followed by the Holling type $I$ functional response. If the birth rate of prey is small enough, then non-lethal effects parameters have no effects on the system stability at coexistence equilibrium. However, if the birth rate of prey is large enough to support oscillation, then non-lethal effects parameters have great impact on the system stability. System can be stable if non-lethal effects parameters are less than some threshold density, which is discussed in Theorems \eqref{thm_critical_value_of_f} and \eqref{thm_critical_value_of_C}. Moreover, at the coexistence equilibrium, equilibrium density of prey does not depend on any of the non-lethal effects parameters, however, predator density decreases as we increase the parameter associated with cost of predation fear $(f)$. On the other hand, the effect of carry-over parameter $(c)$ is opposite, i.e., predator density increases as we increase the parameter $(c)$, when population are at the coexistence steady state. The above discussion fulfills the third objective given in the introduction section regarding the role of non-lethal effects on system stability at the coexistence equilibrium. \\

We provide the global stability conditions of each equilibria in Theorems \eqref{thm_global_stability_boundary_equi_type-I} and \eqref{thm_global_stability_interior_equi_type-I}. As we increase the birth rate parameter $(r)$, our Model \eqref{model_two_species_type-1} experiences two bifurcations of equilibrium and a Hopf-bifurcation at positive equilibria. Existence of Hopf-bifurcation, its direction and stability at the interior equilibrium are discussed in Theorems \eqref{thm_cond_existence_Hopf-bifurcation} and \eqref{thm_cond_direction-stability_Hopf-bifurcation}. From our analysis, we can see that the unique coexistence equilibrium is globally asymptotically stable if the birth rate of prey is not large enough to support the oscillation. One of the interesting result is that our model system supports oscillation, even for the Holling type $I$ functional response, which can not be observed in classical prey-predator model with Holling type $I$ functional response. Even oscillation cannot be observed in the presence of only growth rate reduction due to predation fear and with type $I$ functional response \citep{wang2016modelling, sasmal2018population}. Therefore, the phenomenon 'paradox of enrichment' \citep{rosenzweig1971paradox, gilpin1972enriched} can be observed in our model. However, both analytical and numerical results suggest that such phenomenon can be ruled out by choosing suitable values of cost of fear and/or carry-over effects. This answers the second and fourth questions listed in the introduction section regarding the global dynamical behavior and the phenomenon 'paradox of enrichment.' \\

In the present study, we split the logistic dynamics into birth - death to incorporate the cost of fear only in the birth rate. However, some theoretical study showed that due to the complexity of ecosystem, cost of fear can affect in many ways, like, it may increase the adult death rate, intra-specific competition, etc. \citep{zanette2011perceived, clinchy2013predator, cresswell2011predation}. Therefore, one may consider the fear and its carry-over effect directly to the logistic growth dynamics, when some experimental evidences are available in future. We consider the simplest type $I$ functional response in the presence of predator population. However, a more complicated functional response, like, Holling type $II$ or $III$ functional response can be a mechanism for generating Allee effect in prey due to their predation satiation properties \citep{Gascoigne2004}. Moreover, two or more Allee effects can occur simultaneously in the same population, and this is known as double or multiple Allee effects \citep{berec2007multiple}. Therefore, it may be interesting to see how multiple Allee effects occur due to fear and its carry-over effects when predator follows Holling type $II$ or $III$ functional response. Furthermore, it may be interesting to study how different dynamics can be observed by considering fear and its carry-over effects for other models, for example, Lotka-Volterra model, Beverton-Holt model etc.

\section{Proofs}\label{sec_proofs}

{\bf Proof of the Theorem \eqref{Theorem_existence_stability_single_species}}
\begin{proof}
The local stability of $x_0$, $x_1$ and $x_2$ for Model \eqref{model_single_species} can be determined by the sign of $$\Upsilon(x_i) = \Theta(x_i) + x\Theta^{'}(x_i),$$ where $\Theta^{'}(x_i) = \frac{rcfy}{(1+cx_i+fy)^2}-d_2$. Now, $\Upsilon(x_0) = \Theta(0) = \frac{r}{1+fy}-d_1$. Thus, the equilibrium $x_0$ is locally asymptotically stable if $r<d_1(1+fy)$. \\

For the first part of the Theorem \eqref{Theorem_existence_stability_single_species}, $\Phi(x)>0, \forall x>0$, and consequently, $\Theta(x)<0, \forall x>0$. As \eqref{model_single_species} is a scalar differential equation, $x_0$ attracts every solution, i.e., $x_0$ is globally asymptotically stable. \\

Moreover, $\Theta(x)$ can be written as $$\Theta(x) = -\frac{\Phi(x)}{1+cx+fy} = \frac{(x_1-x)(x-x_2)}{1+cx+fy}.$$ It is easy to obtain,
\begin{eqnarray}\nonumber
\begin{array}{ccc}
\Theta^{'}(x_1) &=& -\frac{(x_1-x_2)}{1+cx_1+fy} (<0) \mbox{ and } \\
\Theta^{'}(x_2) &=& \frac{(x_1-x_2)}{1+cx_2+fy} (>0).
\end{array}
\end{eqnarray}
for $x_1\neq x_2$ (as $x_1>x_2$ for the case when $\Phi(x) = 0$ has real roots). \\

Thus, $\Upsilon(x_1)<0$ and $\Upsilon(x_2)>0$. Therefore, the equilibrium $x_1$ is always locally asymptotically stable when it exists, whereas the equilibrium $x_2$ is always unstable. \\

For part $2(i)$, $x_0$ is unstable as $r>d_1(1+fy)$. Moreover, $\Theta(x)<0$, $\forall x>x_1$ and $\Theta(x)>0$, $\forall x<x_1$. As equation \eqref{model_single_species} is a scalar differential equation, then $x_1$ attracts every solution in $\mathbb{R}_+$, i.e., $x_1$ is globally asymptotically stable. \\

For part $2(ii)$, also the extinction equilibrium $x_0$ is unstable and the positive equilibrium $x_3$ is locally asymptotically stable (as $\Theta^{'}(x_3)<0$). Here also we can easily show that $\Theta(x)<0$, $\forall x>x_3$, and $\Theta(x)>0$, $\forall x<x_3$, and therefore $x_3$ attracts every solution in $\mathbb{R}_+$. \\

For part $2(iii)$, since $\Delta=0$, then $\Omega_2$ must be positive, i.e., $r<d_1(1+fy)$, which is the local stability condition of $x_0$. Here, $\Phi(x) = 0$ has an identical real positive root, and $\Theta^{'}(x_4) = 0$. Therefore, we can't apply eigenvalue approach for the local stability. However, $\Phi(x)>0$ for all $x>0$ and $x\neq x_4$. Therefore, $\Theta(x)<0$, $\forall x>0$ and $x\neq x_4$. Therefore, $x_0$ is locally asymptotically stable as it attracts every solution in $\mathbb{R}_+$ with $x<x_4$ and $x_4$ is a saddle as it attracts every solution, starting at $x>x_4$, but repels $x<x_4$. \\

For the final part, both the equilibria $x_0$ and $x_1$ are locally asymptotically stable. Here, $\Theta(x)<0$, $\forall x>x_1 \mbox{ and } 0 < x < x_2$, $\Theta(x)>0$, $\forall x_2<x<x_1$. Thus the basin of attraction of the extinction equilibrium $x_0$ is $[0,x_2)$, and for the positive equilibrium $x_1$ is $\mathbb{R}_+\backslash[0,x_2)$.
\end{proof}

{\bf Proof of the Lemma \eqref{thm_positivity_boundedness_two_species_type-I}}
\begin{proof}
As $\frac{dx}{dt}\Big|_{x=0} = 0$ and $\frac{dy}{dt}\Big|_{y=0} = 0$ for any $x\geq0$ and $y\geq0$, then $x=0$ and $y=0$ are invariant manifolds, respectively. Due to the uniqueness of solution the set $\mathbb{R}^2_+$ is positively invariant for the Model \eqref{model_two_species_type-1}. Moreover, $$\frac{dx}{dt}\leq\frac{rx(1+cx)}{1+cx+fy}-d_1x-d_2x^2\leq rx-d_1x-d_2x^2.$$ By the comparison theory we can prove that $$\lim_{t\rightarrow\infty}\sup x(t) \leq \frac{r-d_1}{d_2}.$$ Define $w(t) = x(t)+\frac{1}{\alpha}y(y)$, then
\begin{eqnarray}\nonumber
\begin{array}{ccc}
\frac{dw}{dt} &=& \frac{rx(1+cx)}{1+cx+fy}-d_1x-d_2x^2-\frac{m}{\alpha}y \\
&<& rx-d_1x-d_2x^2-m(w-x) \\
&=& (r-d_1+m)x-d_2x^2-mw.
\end{array}
\end{eqnarray}
Then similar to the proof of the theorem $(2.1)$ in \citep{sasmal2020dynamics}, our results follow.
\end{proof}

{\bf Proof of the Theorem \eqref{thm_local_stability_type-I}}

\begin{proof}
The characteristic equation at the interior equilibrium $E_*$ is given by
\begin{eqnarray}\label{eqn_char_int_type-I}
\begin{array}{ccc}
\lambda^2-\Psi_{11}\lambda + \Psi_{12}\Psi_{21} = 0,
\end{array}
\end{eqnarray}
whose roots will be real negative or complex conjugate with negative real parts if $\Psi_{11}<0$, where
\begin{eqnarray}\nonumber
\begin{array}{ccc}
\Psi_{11} &=& x_*\lz\frac{rcfy_*}{(1+cx_*+fy_*)^2} - d_2\rz \\
\Psi_{12} &=& x_*\lz\frac{rf(1+cx_*)}{(1+cx_*+fy_*)^2}+a\rz (>0)\\
\Psi_{21} &=& a\alpha y_* (>0).
\end{array}
\end{eqnarray}
Now, $$\Psi_{11}<0 \mbox{ iff } \frac{rcfy_*}{(1+cx_*+fy_*)^2} < d_2, \mbox{ i.e., iff } r<\frac{d_2(a\alpha+cm+a\alpha fy_*)^2}{a^2\alpha^2cfy_*}.$$

Therefore, $E_*$ is locally asymptotically stable if $r<\frac{d_2(a\alpha+cm+a\alpha fy_*)^2}{a^2\alpha^2cfy_*}$. The local stability of other two equilibria are similar, and hence not discussed here.\\

Moreover,
\begin{eqnarray}\nonumber
\begin{array}{ccc}
&& \frac{rcfy_*}{(1+cx_*+fy_*)^2} - d_2 < 0, \\
&\Leftrightarrow& d_2a^2\alpha^2f^2y_*^2 + \lz2d_2a\alpha f(a\alpha+cm) - ra^2\alpha^2cf\rz y_* + d_2(a\alpha+cm)^2 > 0,
\end{array}
\end{eqnarray}
which is always true if (for any real positive $y_*$)
\begin{eqnarray}\nonumber
\begin{array}{ccc}
&& \lz2d_2a\alpha f(a\alpha+cm)-ra^2\alpha^2cf\rz^2 - 4d_2^2a^2\alpha^2f^2(a\alpha+cm)^2 < 0, \\
&\Rightarrow& r < \frac{4d_2(a\alpha+cm)}{a\alpha c}.
\end{array}
\end{eqnarray}
\end{proof}

{\bf Proof of the Theorem \eqref{thm_critical_value_of_f}}

\begin{proof}
We have, $\frac{rcfy_*}{(1+cx_*+fy_*)^2} - d_2 < rcfy_*-d_2$ and
\begin{eqnarray}\label{cond_sufficient_stability_interior_type-I}
\begin{array}{ccc}
&& rcfy_*-d_2 < 0, \\
&\Leftrightarrow& frc\lz a\alpha r^2c(a\alpha+cm)-rc(a\alpha+cm)(d_1a\alpha+d_2m)-a\alpha d_2(d_1a\alpha+d_2m)\rz \\
&<& a^2\alpha d_2\lz a\alpha d_2+rc(a\alpha+cm)\rz.
\end{array}
\end{eqnarray}
Thus, the condition \eqref{cond_sufficient_stability_interior_type-I} is a sufficient condition for the local stability of $E_*$. \\

Left hand side of the inequality \eqref{cond_sufficient_stability_interior_type-I} will be positive if
\begin{eqnarray}\label{cond_r_crit}
\begin{array}{ccc}
r > \frac{c(a\alpha+cm)(d_1a\alpha+d_2m)+\sqrt{c(a\alpha+cm)(d_1a\alpha+d_2m)\lz c(a\alpha+cm)(d_1a\alpha+d_2m)+4a^2\alpha^2d_2\rz}}{2a\alpha c(a\alpha+cm)} = r_{c}.
\end{array}
\end{eqnarray}
When $r>r_c$, then the sufficient condition for the local stability of $E_*$ is
\begin{eqnarray}\label{cond_f_crit}
\begin{array}{ccc}
f < \frac{a^2\alpha d_2\lz a\alpha d_2+rc(a\alpha+cm)\rz}{rc\lz a\alpha r^2c(a\alpha+cm)-rc(a\alpha+cm)(d_1a\alpha+d_2m)-a\alpha d_2(d_1a\alpha+d_2m)\rz} = f_{c}.
\end{array}
\end{eqnarray}
\end{proof}

{\bf Proof of the Theorem \eqref{thm_critical_value_of_C}}
\begin{proof}
We have,
\begin{eqnarray}\nonumber
\begin{array}{ccc}
&& \frac{rcfy_*}{(1+cx_*+fy_*)^2} - d_2 < 0, \\
&\Leftrightarrow& d_2a^2\alpha^2f^2y_*^2 + \lz2d_2a\alpha f(a\alpha+cm) - ra^2\alpha^2cf\rz y_* + d_2(a\alpha+cm)^2 > 0,
\end{array}
\end{eqnarray}
a sufficient condition, that the above inequality holds true for any positive real $y_*$, is
\begin{eqnarray}\label{cond_sufficient_stability_interior_type-I_for_c}
\begin{array}{ccc}
&& \lz2d_2a\alpha f(a\alpha+cm)-ra^2\alpha^2cf\rz^2 - 4d_2^2a^2\alpha^2f^2(a\alpha+cm)^2 < 0, \\
&\Leftrightarrow& c\lx ra\alpha-4d_2m\rx < 4d_2a\alpha.
\end{array}
\end{eqnarray}
Thus, the condition $r < \frac{4d_2m}{a\alpha}$ is a sufficient condition for the local stability of $E_*$. \\

Since the left hand side of the inequality \eqref{cond_sufficient_stability_interior_type-I_for_c} is positive if $r>\frac{4d_2m}{a\alpha},$ and the sufficient condition for the local stability of $E_*$ is
\begin{eqnarray}\label{cond_c_crit}
\begin{array}{ccc}
c < \frac{4d_2a\alpha}{ra\alpha-4d_2m} = c_r.
\end{array}
\end{eqnarray}
\end{proof}

{\bf Proof of the Theorem \eqref{thm_global_stability_boundary_equi_type-I}}

\begin{proof}
The global stability of $E_0$ in $r\in(0,d_1)$ follows from lemma \eqref{thm_positivity_boundedness_two_species_type-I} and Theorem \eqref{thm_local_stability_type-I}. Moreover, when $r\in\lx d_1,d_1+\frac{d_2m}{a\alpha}\rx$, there exists only two equilibria $E_0$ and $E_1$ in $\mathbb{R}^2_+$, and hence there can not be any periodic orbit in $\mathbb{R}^2_+$, which implies that every solution will converge to any one of $E_0$ and $E_1$. However, when $r\in\lx d_1,d_1+\frac{d_2m}{a\alpha}\rx$ then $E_0$ is a saddle (repelling) and every solution will approach to $E_1$. Therefore, from local stability condition, $E_1$ exists and is globally asymptotically stable if $r\in\lx d_1,d_1+\frac{d_2m}{a\alpha}\rx$.
\end{proof}

{\bf Proof of the Theorem \eqref{thm_global_stability_interior_equi_type-I}}

\begin{proof}
A sufficient condition for the local stability of the positive equilibrium is $r<\frac{4d_2(a\alpha+cm)}{a\alpha c}$. Since, $\frac{4d_2}{c} < \frac{4d_2(a\alpha+cm)}{a\alpha c}$, from the previous Theorems \eqref{thm_local_stability_type-I} and \eqref{thm_global_stability_boundary_equi_type-I}, to show the global stability of $E_*$, it is sufficient to prove that for $r\in\lx d_1+\frac{d_2m}{a\alpha}, \frac{4d_2}{c}\rx$, there is no periodic orbit in $\{(x,y)|x>0,y>0\}$. \\

By taking the Dulac function $D(x,y) = \frac{1}{xy}$ for the System \eqref{model_two_species_type-1}, we have
\begin{eqnarray}\nonumber
\begin{array}{ccc}
\mbox{div}\Big|_{\lx D\frac{dx}{dt},D\frac{dy}{dt}\rx} &=& \frac{\partial}{\partial x}\lx D(x,y)\frac{dx}{dt}(x,y)\rx + \frac{\partial}{\partial y}\lx D(x,y)\frac{dy}{dt}(x,y)\rx \\
&=& \frac{1}{y}\lz\frac{rcfy}{(1+cx+fy)^2}-d_2\rz \\
&=& \frac{1}{y(1+cx+fy)^2}\lz-d_2f^2y^2 + \{rcf - 2d_2f(1+cx)\}y - d_2(1+cx)^2\rz.
\end{array}
\end{eqnarray}
It is easy to see that $\lz rcf-2d_2f(1+cx)\rz^2 - 4d_2^2f^2(1+cx)^2 < 0$ if $r<\frac{4d_2}{c}$. \\

Therefore, $\mbox{div}\Big|_{\lx D\frac{dx}{dt},D\frac{dy}{dt}\rx}<0$ in $\{(x,y)|x>0,y>0\}$ if $r<\frac{4d_2}{c}$. Then by the Dulac-Bendixson theorem \citep{perko2013differential}, there is no periodic orbit in $\{(x,y)|x>0,y>0\}$ for System \eqref{model_two_species_type-1} if $r<\frac{4d_2}{c}$. Moreover, $E_*$ is the only stable equilibrium in $\{(x,y)|x>0,y>0\}$ if $d_1+\frac{d_2m}{a\alpha}<r<\frac{4d_2}{c}$. Hence every positive solution will tend to $E_*$.
\end{proof}

{\bf Proof of the Theorem \eqref{thm_cond_direction-stability_Hopf-bifurcation}}

\begin{proof}
To find the stability and direction of Hopf-bifurcation, we calculate the $1$st Lyapunov coefficient. Let $u=x-x_*$ and $v=y-y_*$, then the System \eqref{model_two_species_type-1} becomes
\begin{eqnarray}\nonumber
\begin{array}{ccc}
\frac{du}{dt} &=& \frac{r(u+x_*)(1+c(u+x_*))}{1+c(u+x_*)+f(v+y_*)} - d_1(u+x_*) - d_2(u+x_*)^2 - a(u+x_*)(v+y_*) := g(u,v) \\
\frac{dv}{dt} &=& a\alpha(u+x_*)(v+y_*) - m(v+y_*) := h(u,v).
\end{array}
\end{eqnarray}
Now, considering the Taylor's series expansion at $(u,v)=(0,0)$ up to 3rd order, we have
\begin{eqnarray}\label{eqn1_lyapunov}
\begin{array}{ccc}
\frac{du}{dt} &=& g_uu + g_vv + g_1(u,v), \\
\frac{dv}{dt} &=& h_uu + h_vv + h_1(u,v),
\end{array}
\end{eqnarray}
$g_1(u,v)$ and $h_1(u,v)$ are the higher order terms of $u$ and $v$, given by
\begin{eqnarray}\nonumber
\begin{array}{ccc}
g_1(u,v) &=& g_{uu}u^2 + g_{uv}uv + g_{vv}v^2 + g_{uuu}u^3 + g_{uuv}u^2v + g_{uvv}uv^2 + g_{vvv}v^3, \\
h_1(u,v) &=& h_{uu}u^2 + h_{uv}uv + h_{vv}v^2 + h_{uuu}u^3 + h_{uuv}u^2v + h_{uvv}uv^2 + h_{vvv}v^3,
\end{array}
\end{eqnarray}
where
\begin{eqnarray}\nonumber
\begin{array}{ccc}
g_u &=& - d_2x_* + \frac{rcfx_*y_*}{(1+cx_*+fy_*)^2}, \mbox{~~~~} g_v = -\lz ax_* + \frac{rfx_*(1+cx_*)}{(1+cx_*+fy_*)^2}\rz, \mbox{~~~~} g_{uu} = -d_2 - \frac{rcfy_*(-1+cx_*-fy_*)}{(1+cx_*+fy_*)^3}, \\\\
g_{uv} &=& \frac{rcfx_*(1+cx_*-fy_*)}{(1+cx_*+fy_*)^3}, \mbox{~~~~} g_{vv} = -\frac{2rx_*(1+cx_*)f^2}{(1+cx_*+fy_*)^3}, \mbox{~~~~} g_{uuu} = \frac{2rc^2fy_*(-2+cx_*-2fy_*)}{(1+cx_*+fy_*)^4}, \\\\
g_{uuv} &=& -\frac{rcf(-1+c^2x_*^2+f^2y_*^2-4cx_*fy_*)}{(1+cx_*+fy_*)^4}, \mbox{~~~~} g_{uvv} = -\frac{2rcf^2x_*(2+2cx_*-fy_*)}{(1+cx_*+fy_*)^4}, \mbox{~~~~} g_{vvv} = \frac{6rx_*(1+cx_*)f^3}{(1+cx_*+fy_*)^4},
\end{array}
\end{eqnarray}
and
\begin{eqnarray}\nonumber
\begin{array}{ccc}
h_u &=& a\alpha y_*, \mbox{~~} h_v = 0, \mbox{~~} h_{uu} = 0, \mbox{~~} h_{uv} = a\alpha, \mbox{~~} h_{vv} = 0, \mbox{~~} h_{uuu} = 0, \mbox{~~} h_{uuv} = 0, \mbox{~~} h_{uvv} = 0, \mbox{~~} h_{vvv} = 0.
\end{array}
\end{eqnarray}

Here all the partial derivatives are calculated at the bifurcation point, i.e., $(u,v) = (0,0)$. Thus system \eqref{eqn1_lyapunov} can be written as $$\dot{U} = \left[
\begin{array}{cc}
g_u & g_v \\
h_u & h_v
    \end{array}
\right]U + F(u),$$
where $U=(u,v)^T$ and {\small{$F = \lx f_1(u,v),g_1(u,v)\rx^T = \lx g_{uu}u^2 + g_{uv}uv + g_{vv}v^2 + g_{uuu}u^3 + g_{uuv}u^2v + g_{uvv}uv^2 + g_{vvv}v^3, h_{uv}uv\rx^T$.}} \\

Now, Hopf-bifurcation occurs when $g_u = 0$, i.e., at the Hopf-bifurcation point, the eigenvalue will be purely imaginary, which is given by $i\omega$, where $\omega=\sqrt{-g_vh_u}$. Eigenvector corresponding to this eigenvalue $i\omega$ is given by $\bar{v}=\lx g_v,i\omega\rx^T$. Now, we define $Q = \lx Re(\bar{v}), -Im(\bar{v})\rx = \left[
\begin{array}{cc}
g_v & 0 \\
0 & i\omega
    \end{array}
\right]$.
Now, let $U=QZ$ or $Z=Q^{-1}U$, where $Z=(z_1,z_2)^T$. Therefore, under this transformation, the system is reduced to
$$\dot{Z} = \lx Q^{-1}\left[
\begin{array}{cc}
g_u & g_v \\
h_u & h_v
    \end{array}
\right]Q\rx Z + Q^{-1}F(QZ).$$
 This can be written as
$$\left[
\begin{array}{cc}
\dot{z}_1 \\
\dot{z}_2
    \end{array}
\right] = \left[
\begin{array}{cc}
0 & -\omega \\
\omega & 0
    \end{array}
\right]\left[
\begin{array}{cc}
z_1 \\
z_2
    \end{array}
\right] + \left[
\begin{array}{cc}
F_1(z_1,z_2) \\
F_2(z_1,z_2)
    \end{array}
\right],$$

where $F_1(z_1,z_2)$ and $F_2(z_1,z_2)$ are given by

\begin{eqnarray}\nonumber
\begin{array}{ccc}
F_1(z_1,z_2) &=& \frac{1}{g_v}\lz g_{uu}g_v^2z_1^2 - \omega g_{uv}g_vz_1z_2 - g_vh_ug_{vv}z_2^2 + g_{uuu}g_v^3z_1^3 - \omega g_{uuv}g_v^2z_1^2z_2 - g_vh_ug_{uvv}g_vz_1z_2^2 - g_vh_u\omega g_{vvv}z_2^3\rz \\
F_2(z_1,z_2) &=& g_vh_{uv}z_1z_2.
\end{array}
\end{eqnarray}

The direction of Hopf-bifurcation is determined by the sign of the $1$st Lyapunov coefficient, which is given by

\begin{eqnarray}\nonumber
\begin{array}{ccc}
L &:=& \frac{1}{16}\lz\frac{\partial^3F_1}{\partial z_1^3} + \frac{\partial^3F_1}{\partial z_1\partial^2z_2} + \frac{\partial^3F_2}{\partial^2z_1\partial z_2} + \frac{\partial^3F_2}{\partial^3z_2}\rz \\
&+& \frac{1}{16\omega}\lz\frac{\partial^2F_1}{\partial z_1\partial z_2}\lx\frac{\partial^2F_1}{\partial^2z_1}+\frac{\partial^2F_1}{\partial^2z_2}\rx - \frac{\partial^2F_2}{\partial z_1\partial z_2}\lx\frac{\partial^2F_2}{\partial^2z_1}+\frac{\partial^2F_2}{\partial^2z_2}\rx - \frac{\partial^2F_1}{\partial^2z_1}\frac{\partial^2F_2}{\partial^2z_1} + \frac{\partial^2F_1}{\partial^2z_2}\frac{\partial^2F_2}{\partial^2z_2}\rz.
\end{array}
\end{eqnarray}

We use the maple software to simplify the expression of $L$, which is given as follows:

\begin{eqnarray}\nonumber
\begin{array}{ccc}
L &:=& 3g_v^2g_{uuu}\lz(1-g_{uv}z_1)+2z_1(g_{uvv}z_2w-g_{uuv}z_1g_v)\rz + 3wg_{vvv}z_2\lz2w(h_ug_{uvv}z_2-wg_{uuv}z_1)+g_{uv}h_u\rz \\
&+& g_{uvv}w\lz w(1-g_{uv}z_1)+2z_2(g_vg_{uu}-h_ug_{vv})+2z_2w(wg_{uvv}z_1-g_vz_2g_{uuv})\rz \\
&+& 2g_{uuv}z_1\lz g_v^2(wg_{uuv}z_2-g_{uu})-w^2(g_{vv}+g_vz_1g_{uvv})\rz + g_{uv}(h_ug_{vv}+wg_vg_{uuv}z_2-g_vg_{uu}).
\end{array}
\end{eqnarray}

Now by \citep{perko2013differential}, Hopf-bifurcation is supercritical if $L<0$ and it is subcritical if $L>0$.
\end{proof}

\section*{Acknowledgement}
YT's research is supported by Aoyama Gakuin University research grant ``Ongoing Research Support" and Japan Society for the Promotion of Science ``Grand-in-Aid 20K03755."

\section*{References}
\bibliographystyle{elsarticle-harv}
\bibliography{sasmalbib_fear_carry-over}

\end{document}